\documentclass[aps,prd,a4paper,twocolumn,notitlepage,nofootinbib,groupedaddress,showpacs,showkeys]{revtex4-1}
\pdfoutput=1
\usepackage{amsmath}
\usepackage{amssymb}
\usepackage{amsfonts}
\usepackage{latexsym}
\usepackage{txfonts}
\usepackage{hhline}
\usepackage{graphicx}
\usepackage[colorlinks=true]{hyperref}
\usepackage{color}
\usepackage{appendix}
\usepackage{acronym}
\usepackage{dcolumn}   
\usepackage{url}
\usepackage{framed}
\usepackage{cleveref}
\usepackage{caption}
\usepackage{subfigure}

\newcommand{\dcc}{LIGO-P1500228}

\begin{document}

\title{A classifier for gravitational-wave inspiral signals in non-ideal single-detector data}

\author{S.~J.~Kapadia}~\email{kapadia@uwm.edu}
\affiliation{Department of Physics, University of Arkansas, Fayetteville, AR 72701, USA}
\affiliation{Albert-Einstein-Institut (Max-Planck-Institut f{\"ur} Gravitationsphysik), Callinstr.~38, Hannover, Germany}
\affiliation{Center for Gravitation, Cosmology, and Astrophysics, University of Wisconsin-Milwaukee, Milwaukee, WI 53201, USA}
\author{T.~Dent}~\email{thomas.dent@aei.mpg.de}
\affiliation{Albert-Einstein-Institut (Max-Planck-Institut f{\"ur} Gravitationsphysik), Callinstr.~38, Hannover, Germany and Leibniz-Universit{\"a}t Hannover, Germany}
\author{T.~\surname{Dal Canton}}~\email{tito.dalcanton@aei.mpg.de}
\affiliation{Albert-Einstein-Institut (Max-Planck-Institut f{\"ur} Gravitationsphysik), Callinstr.~38, Hannover, Germany}
\affiliation{NASA Postdoctoral Program Fellow, Goddard Space Flight
Center, Greenbelt, MD 20771, USA}
\date{\today, \dcc}

\begin{abstract}
\noindent
We describe a multivariate classifier for candidate events in a templated search for 
gravitational-wave (GW) inspiral signals from neutron-star--black-hole (NS-BH) binaries, in
data from ground-based detectors where sensitivity is limited by non-Gaussian noise 
transients.  The standard signal-to-noise ratio (SNR) and chi-squared test for inspiral 
searches use only properties of a single matched filter at the time of an event; instead, 
we propose a classifier using features derived from a bank of inspiral templates around 
the time of each event, and also from a search using approximate sine-Gaussian templates.  
The classifier thus extracts 
additional information from strain data to discriminate inspiral signals from noise 
transients.  We evaluate a Random Forest classifier on a set of single-detector events 
obtained from realistic simulated advanced LIGO data, using simulated NS-BH signals added to 
the data.  The new classifier detects a factor of 1.5 -- 2 more signals at low false positive
rates as compared to the standard `re-weighted SNR' statistic, and does not require the 
chi-squared test to be computed. 
Conversely, if only the SNR and chi-squared values of single-detector events are 
available, Random Forest classification performs nearly identically to the re-weighted 
SNR. 
\end{abstract}
\maketitle

\section{Introduction}
\subsection{Motivation}
\noindent
The epoch of gravitational wave astronomy has begun with the unambiguous detection of GW
signals from massive merging binary black hole (BBH) 
systems~\cite{Detection,CBCCompanion,DetcharCompanion,PECompanion,BoxingDay,O1BBH,GW170104} 
in data from the Advanced LIGO interferometers~\cite{DetectorCompanion}.  
Already, though, the non-detection of binary neutron-star (BNS) and neutron star-black
hole (NS-BH) binaries in the first Advanced LIGO observing run~\cite{O1BNSNSBH} provides  
motivation to further develop 
search methods for coalescing binaries (CBC) in order to fully realize the science 
potential of the Advanced detector network~\cite{TheLIGOScientific:2014jea,
TheVirgo:2014hva}.  
Moreover, since it appears that elucidating the origin of merging BBH systems may require 
some tens of detections (see e.g.~\cite{Stevenson_misalign}), it is also desirable to 
increase the sensitivity of searches to relatively weak CBC signals, which should be more
numerous than high-SNR detections.  

Consider a signal comparable to the candidate event LVT151012, which has a network 
SNR of $\sim 9.7$, consistent with a massive BBH merger at redshift $\sim 0.2$ if 
astrophysical, but is assigned a false alarm rate of $0.4$ per year in the advanced 
LIGO-Virgo search pipeline of~\cite{CBCCompanion,O1BBH,Usman}.  (This false 
alarm rate is the expected number of noise events with a higher ranking than LVT151012 in 
the given search pipeline, per year of data searched.)  We could not confidently rule out 
that an event with comparable SNR was due to noise with current methods.  Noise events 
louder than such relatively weak signals
are still dominated by transient detector artefacts (`glitches'), which are generally 
suppressed by the standard chi-squared test~\cite{Brucechisq,FindChirp} but not 
eliminated.  If new methods are able to further reduce the contribution of non-Gaussian 
artefacts to the noise background, 
the search sensitivity to weak signals could be significantly increased, as an improved
pipeline would assign such events lower false alarm rates relative to current methods,  
corresponding to a higher probability of astrophysical origin via the analysis of~\cite{RatesCompanion,RateSuppl}.

\subsection{Detection statistics and followup for inspiral events}
Searches for GW from inspiraling compact binary sources in data from 
ground-based detectors~\cite{ihope,S5Year1,S51218,S5LV,S6Lowmass} have so far relied on 
empirical methods to suppress the effects of non-Gaussian noise transients (`glitches') 
\cite{Blackburn:2008ah, S5DQ, S6DC}.  These transients give rise to a background 
distribution of `triggers' (maxima of the matched filter SNR time series~\cite{FindChirp}) with SNRs up to $10^2$--$10^3$, whereas Gaussian noise would, for 
typical search parameters, produce a maximum network SNR of order 10 or less.  
Without any steps to exclude or down-rank very high noise SNRs, the search sensitivity 
for astrophysical signals would be reduced by orders of magnitude~\cite{ihope}.

Various signal consistency tests for loud triggers in binary inspiral searches have been
considered~\cite{Brucechisq,Rodriguez:2008kt,Babak2005,Hannathesis,Harry:2010fr,gstlal2016}; the time-frequency 
$\chi^2$ test described in~\cite{Brucechisq} has been widely employed, due to its
relative simplicity and effectiveness over a range of different epochs of data and 
signal parameters.  In `all-sky' searches (those without a restriction on the times
and sky directions searched) candidate events have been ranked by a simple algebraic 
function of the matched filter SNR $\rho$ and $\chi^2$ of single-detector
triggers (`effective SNR' or `re-weighted SNR')~\cite{ihope,S6Lowmass} which pass a 
consistency test between their arrival time and mass parameters~\cite{ethinca}. 

\begin{figure}
\centering
\includegraphics[trim=13mm 77mm 10mm 75mm, clip, width=0.95\columnwidth]{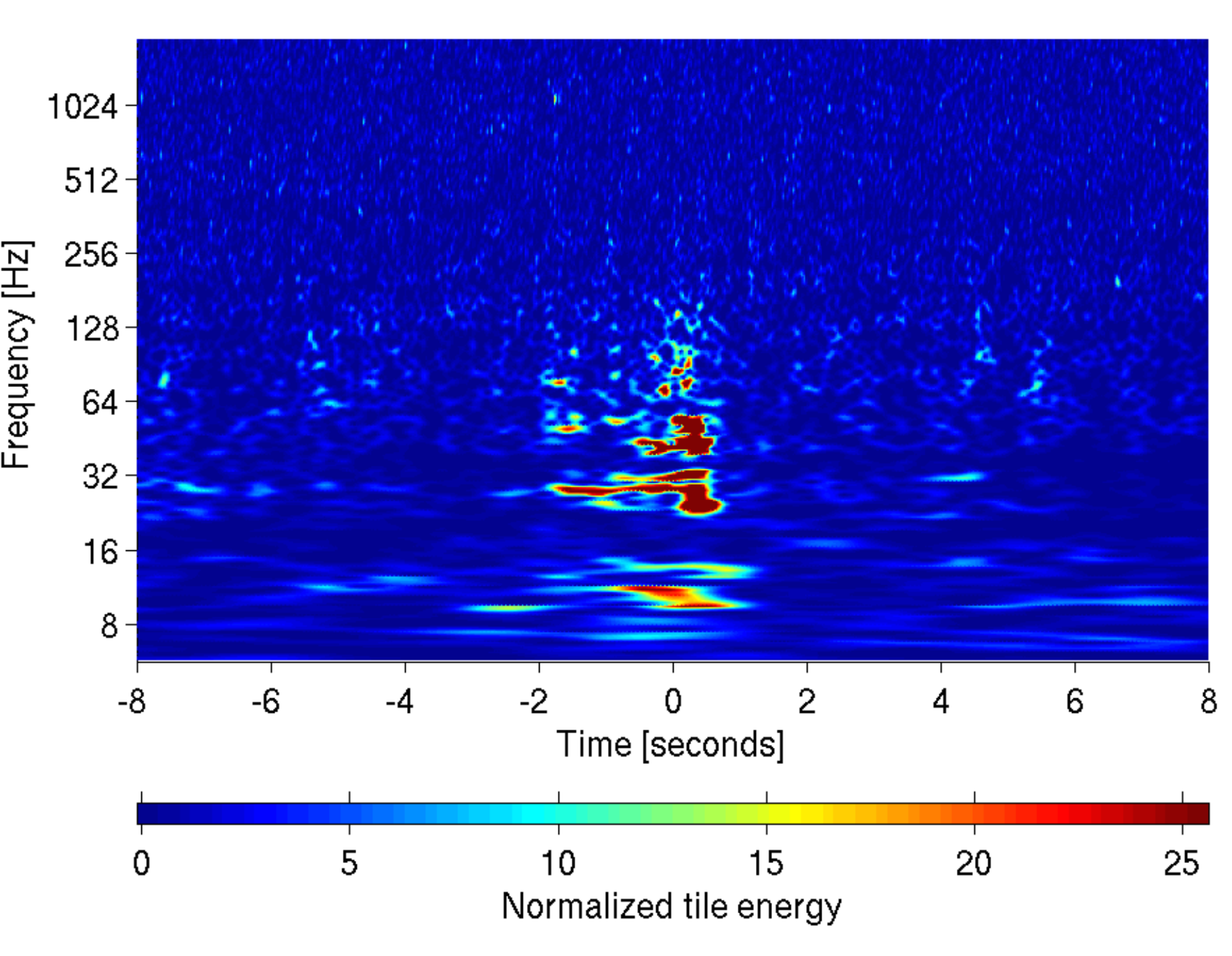}
\caption{Diagnostic whitened spectrogram (`Omegagram') from a decomposition of LIGO 
strain data in an approximate sine-Gaussian basis, showing excess noise power (`tile 
energy') 
over for a few seconds of searched data. 
Such non-Gaussian noise gives rise to 
some of the highest-ranked events in a matched filter search for inspiral signals 
where the well-known time-frequency $\chi^2$ test is employed.\label{fig:omega_scan_glitch}}
\end{figure}

In principle the loudest search events, i.e.\ those with highest combined re-weighted 
SNR $\hat{\rho}$, are the most likely to indicate GW 
signals~\cite{ihope,S6Lowmass}.
Conversely, the loudest events generated by unphysical relative time-shifts of 
detectors at different locations, used to estimate the noise background of the search, 
are those which restrict the sensitivity of the search at low false alarm rate. 
However, when examining the properties of these loudest events in detail 
\cite{Gouaty:2008gn}, we often find that even triggers with high $\hat{\rho}$ appear 
to be caused by loud glitches, or occur in times of 
sub-optimal data quality as shown by the presence of excess noise over periods of 
$~$seconds.  Such excess noise can be diagnosed by the presence of many high-energy 
tiles when decomposing the strain data in an approximate sine-Gaussian basis~\cite{ShourovThesis};
a `Q-scan' or `Omega scan' diagnostic output for a few seconds of poor quality data from 
the LIGO S6 science run is shown in Fig.~\ref{fig:omega_scan_glitch}.  In contrast, 
an inspiraling binary signal in Gaussian noise viewed in this basis would either 
have no loud tiles at all, or if the signal was strong, the loud tiles would trace a 
clear `chirp' trajectory in time and frequency, as in Fig.~1 of~\cite{Detection} 
and (for simulated signals) in Fig.~6 of~\cite{Gouaty:2008gn} and the LIGO-Virgo S6/VSR3 
blind injection data release~\cite{GW100916spec}.

Furthermore, when examining the triggers from the inspiral search around the time of
a candidate (plots known as `mini-followup'), we often find that the loudest events show 
large numbers of high-SNR triggers spread over several seconds, whereas for an 
inspiral signal in Gaussian noise we expect a small number of high-SNR triggers
localized within a fraction of a second of the loud event and with similar chirp mass, 
defined as  
\begin{equation} \label{eq:mchirp}
\mathcal{M} = (m_1m_2)^{3/5}(m_1+m_2)^{-1/5},
\end{equation}
where $m_{1,2}$ are the binary component masses.  See the right-hand plots in 
Figures~\ref{NoiseFollowup} and~\ref{SignalFollowup} respectively for examples of 
`glitchy' vs.\ 'clean' sets of triggers around a loud event. 
These two plots can easily be distinguished although the highest-SNR trigger has similar 
properties in both.  Thus, the strain data around the time of a noise trigger of high $
\hat{\rho}$ will likely contain more information than the $\chi^2$ statistic, which   
uses only the samples of the matched filter integral for 
that trigger, divided into frequency bins~\cite{Brucechisq}.

\subsection{Outline of classifier method} 
Our objective here is to identify additional information available in the strain data 
and employ it effectively in distinguishing triggers due to binary inspiral-merger 
signals from those caused by noise transients.  To do so we will construct a ranking 
statistic that uses several independent, or partially correlated, pieces of information 
(`features') extracted from the data around each inspiral search trigger. 
Some of these features consist of information already available in the output of the 
inspiral matched filter search, namely the coalescence times, SNRs and parameters of 
other triggers close to the event under consideration.  

Our classification information also includes the output of another analysis algorithm, 
\emph{omicron}~\cite{OmicronRobinet},
which responds to transient signals or artefacts in a very different way from the  
inspiral matched filter search.  
Omicron is an adaptation of the Q-pipeline 
and Omega search methods \cite{ShourovThesis,S5BurstPaper} which aim to detect transient 
events with excess power localized in time and frequency, without placing strong 
constraints on the morphology of such events.  
Omega/omicron have 
also been applied to single-interferometer data 
for characterization of non-Gaussian noise transients in LIGO and Virgo data 
\cite{S6DC,VirgoDC}; our use of 
omicron has a similar motivation, as we are 
seeking to down-rank times where strain data contains high-amplitude noise 
artifacts. 
A set of omicron triggers may contain information on such noise events which is 
at least partly independent of that contained in inspiral matched filter triggers.

To combine features derived from omicron triggers with those from the inspiral search 
triggers, we implement a Random Forest multivariate classifier~\cite{RandomForest} trained 
on both noise triggers derived from realistic simulated early Advanced LIGO data, and on 
simulated neutron-star black-hole (NS-BH) binary signals added to such data.  We then use 
independent trigger samples to evaluate how efficiently the classifier sorts noise from 
signal events. 

\subsection{Relation to previous work}
Multivariate classification methods to separate transient GW signals from non-Gaussian 
artefacts have been proposed and implemented in the context of templated searches for 
inspiral-merger-ringdown (IMR) signals from stellar-mass or intermediate mass binary black
holes~\cite{Baker:2014eba,Hodge:2014yca}, and in a search for weakly-modelled GW burst 
signals associated with external high-energy electromagnetic triggers~\cite{Adams:2013pna}. 
The possible improvement in efficiency of a coherent templated search for inspiral GW 
signals associated with gamma-ray bursts (GRB)~\cite{Harry:2010fr} due to use of a neural 
network was also investigated in~\cite{Kim:2014nba}.  Our work differs from these methods
in two main respects.  First, in contrast to previous work which uses properties of events
derived from multi-detector analysis, 
we show that 
multivariate classification can be effective even using only the data stream from a single 
detector.  Our approach may then be useful to `modularize' the classification problem for 
a multi-detector network, splitting the calculation into parts which depend on the noise 
properties in each separate detector and parts which depend on the joint properties 
of an event compared in several detectors. 

Second, we build our classification information from the outputs of two entirely 
independent analyses applied to a given data stream, effectively viewing the data in two 
different ways, which allows us to extract information which would not be available when
using the output of a single search pipeline alone.  In the case we consider, the omicron
(burst) analysis acts as a diagnostic for non-Gaussian noise
similar to a followup by visual inspection of the data.  We also use information 
from times surrounding a candidate trigger, rather than only at the peak of the likelihood; 
this information helps to diagnose the state of the detector, uncovering times of worse 
quality data where frequent glitches occur.  


\paragraph*{Overview of the paper}  We will proceed in Section~\ref{sec:GeneralFramework} 
by setting out the problem to be addressed in classifying inspiral search events (triggers) 
and defining a figure of merit for the outcome of the classification.  We give a brief 
introduction to the Random Forest method of multivariate classification in 
Section~\ref{sec:RF}, then describe its application to the case of a templated search for 
inspiral GW signals in Section~\ref{sec:features}.  The tuning of classifier parameters and 
the results of evaluating the classifier on simulated non-ideal early Advanced era detector 
data are given in Section~\ref{sec:results}, with Section~\ref{sec:concl} giving 
conclusions and discussion of possible further steps.

\section{Statistics of event classification}\label{sec:GeneralFramework}
The detection of transient gravitational-wave signals in data from interferometric detectors 
such as Advanced LIGO and Virgo~\cite{TheLIGOScientific:2014jea,TheVirgo:2014hva}
is a statistical classification problem.  Given the expected rates of occurrence of 
source events~\cite{Abadie:2010cf}, most time samples
in the detector outputs will contain either no signal, or a signal that is so weak as to be 
indistinguishable from zero by any means in the presence of detector noise.  The task is then 
to sort the few time samples that are likely to contain an identifiable signal, and thus 
contain information on the nature of a GW source, from the overwhelming majority of 
noise-only samples. 

This sorting is done by considering short time periods (such that each is extremely
unlikely to contain more than $1$ signal) and assigning each period labelled by $i$ a real 
number $\Lambda_i$, where large $\Lambda$ indicates a greater likelihood that a signal is 
present and small $\Lambda$ indicates a smaller likelihood of signal, i.e.\ a greater 
likelihood that time $i$ contains only noise.  Times when $\Lambda$ exceeds a threshold 
$\Lambda_t$ are considered as candidate signals. 

Due to the presence of random (unpredictable) noise in the detector outputs, this 
classification can never be perfectly reliable.  Even with a high threshold $\Lambda_t$, the 
false positive rate (FPR) $p_f(\Lambda_t) \equiv p(\Lambda_i>\Lambda_t|N)$, where $N$ 
indicates the hypothesis of data containing noise only, can never be reduced to zero.  
Conversely, raising the threshold will cause the detection probability $p_d(\Lambda_t) \equiv 
p(\Lambda_i>\Lambda_t|S)$ to decrease, where $S$ indicates the hypothesis of data containing
noise plus a signal of non-negligible amplitude.  
In addition, if the signals have a distribution of amplitudes, the weaker signals are less 
likely to produce a $\Lambda$ value above any specific threshold.  A classification method 
mapping the data at time $i$ to a $\Lambda_i$ value is optimized by maximizing $p_d$, 
marginalized over a given distribution of signal parameters \cite{Searle:2008jv}, at fixed 
$p_f$.  In practice the threshold $\Lambda_t$ may be adjusted to obtain a desired $p_f$ 
value. 

An optimal search statistic will then be given by the likelihood ratio $\Lambda_{i,{\rm opt}} 
\equiv p(d_i|S)/p(d_i|N)$: in words, the relative likelihood that the data at 
time $i$ is caused by a signal plus noise, versus by noise only.\footnote{Any monotonic 
function of $\Lambda_{\rm opt}$ will result in the same relative ranking, and is 
thus also optimal.}  Our task is then, if possible, to evaluate $\Lambda_{i,{\rm opt}}$ for 
the actual noise seen at detector outputs and for the desired signals; if this is not 
possible from first principles, then to obtain a good approximation to it. 
 
The matched filter is an optimal method for a signal of known form in stationary noise, and 
may straightforwardly be extended to post-Newtonian signal waveforms from quasi-circular 
non-precessing binary systems for which the coalescence time, phase and amplitude are 
unknown~\cite{Sathyaprakash:1991mt,FinnChernoff:1993,FindChirp}.  Briefly, each signal 
template is correlated with the strain data in Fourier domain, weighting by the inverse of 
the detector noise power spectrum (power spectral density, PSD).  The resulting time series, 
normalized such that the template's correlation 
with itself is unity, is maximized over complex phase to obtain a real SNR $\rho(t)$, 
which is then maximized over time (over periods of typically a few seconds) to obtain 
triggers.  The trigger SNR $\rho_i$ is, to a good approximation, a monotonic function 
of $\Lambda_i$, and is thus suitable as a detection statistic. 


The above is only true if the noise is indeed Gaussian and stationary; however, real 
detector strain 
shows a large number of transients well localized in time -- ``glitches'' -- caused by more 
or less well-understood instrumental or environmental effects; see e.g.~\cite{Abbott:2003pj,
Blackburn:2008ah,S6DC,VirgoDC}.  Although many such transient artefacts may be readily 
identified and removed from searches~\cite{S5DQ}, large numbers remain unexplained, 
particularly with relatively low amplitude (SNRs of order 10-20; though such values are 
still strongly inconsistent with Gaussian noise).  The presence of non-stationary transients 
invalidates the matched filter SNR as a detection statistic and ranking events by $\rho_i$ 
then leads to a vast loss of sensitivity to signals compared to a search in Gaussian noise 
with comparable PSD~\cite{Usman}.

Hence a different strategy must be adopted.  Searches for inspiraling binary signals in 
LIGO-Virgo data have implicitly assumed that the majority of time samples are well
modelled by stationary, Gaussian noise, with a small number of times affected by glitches.  
This motivates 
calculating additional quantities besides $\rho$ which will indicate the presence of a 
glitch, then vetoing (removing) or down-ranking affected times.  A widely used test
is the time-frequency chi-squared~\cite{Brucechisq} which splits the template 
waveform into several frequency bands and checks whether the matched filter output for each
one at the time of the supposed signal is consistent with expected amplitude and 
phase.  
High-amplitude glitches have large $\chi^2$ values relative to the expectation in Gaussian 
noise, while signals which are well matched to a given template have small $\chi^2$ 
(approximately $1$ per degree of freedom).  Recent searches have used the ``re-weighted SNR''
$\hat{\rho}(\rho,\chi^2)$ of single-detector triggers as a ranking 
statistic~\cite{S6Lowmass,Babak:2012zx,CBCCompanion,Usman}:
\begin{equation}
\hat{\rho} =
  \begin{cases}
    \rho &\textrm{for } \chi^2 \leq n_{\rm dof}\\
    \rho \left[\left(1+(\frac{\chi^2}{n_{\rm dof}})^3\right)/2\right]^{-\frac{1}{6}} &\textrm{for } \chi^2 > n_{\rm dof} ,
  \end{cases}
\label{eq:newSNR}
\end{equation}
where $n_\mathrm{dof} = 2p-2$ is the number of degrees of freedom of the $\chi^2$ test with
$p$ frequency bands.  (For multi-detector events, the ranking statistic is taken to be the 
quadrature sum of $\hat{\rho}$ values over detectors.) 

We would like to generalize such a ranking statistic to a function
of several pieces of information available for each binary merger search trigger, 
called `features'.  For instance, each trigger has an SNR value, a template chirp 
mass $\mathcal{M}$, etc..  We notate each trigger's features as a $p$-dimensional vector 
${\bf x}$; triggers resulting from detector noise are written as ${\bf x}_n$, those arising
from (simulated or real) signal added to noise as ${\bf x}_s$.  We write the total number of 
noise and signal triggers as $N_n$ and $N_s$, respectively.   
 
Given a classification method which assigns a trigger with features ${\bf x}$ a likelihood
of belonging to either ${\bf x}_n$ or ${\bf x}_s$, we require a method to assess the 
performance of the classifier.  Typically, a classifier needs to ``train'' itself on a set of 
candidate events where the status of each ${\bf x}$ as noise or signal is already known.  
Given the low rate of detected GW in existing data and the trigger generation threshold
adopted in this analysis, 
the great majority of triggers will be due to noise.  To 
train a classifier we also require a large number of simulated GW signals, 
typically via injecting (adding) the signal strain to the detector data to produce a set of 
candidate events.


In principle, one could train the classifier on the entire set of simulated triggers at hand,
and test it on the same set of triggers. This method however is susceptible to overfitting, 
making it difficult to trust the classifier's predictions on trigger sets it has not trained
on. In order to circumvent this problem, cross-validation is employed: 
a portion of the trigger set is kept aside for testing, while the 
remaining triggers are used for training. A more sophisticated version, 
stratified K-fold cross validation, splits the data into $K$ partitions or 
``folds'', where $K$ is an integer greater than unity. Stratification ensures that each 
partition is well balanced, and not skewed in favour of one or other class.  
The classifier then trains itself on $K-1$ folds, the remaining fold having been kept aside 
for testing. This process is repeated $K$ times, with each repetition using a different 
fold for testing and correspondingly a different set of folds for training.

There are two kinds of classification models: discrete and probabilistic. The discrete 
classifier directly labels a candidate event ${\bf x}$ in a test set as either ${\bf x}_n$ or 
${\bf x}_s$. The probabilistic classifier assigns a score $\hat{p}$ to a trigger, the score 
being a continuous variable.  Typically, though not necessarily, $\hat{p}$ is an estimate of 
the probability that a trigger is a signal event. 
The Random Forest multivariate classifier used in this project outputs such an estimate, 
which may then be used to predict whether the trigger is associated with a noise or signal 
event by setting a threshold value $\hat{p}_t$ for the classifier score.   

For each prediction of a trigger's category, there are four possible cases.  
These may be 
summarized by a so-called confusion matrix, with the four elements ``True 
Positive'', ``False Positive'', ``False Dismissal'', ``True Dismissal''.  True Positive (True 
Dismissal) corresponds to a signal (noise) trigger accurately labelled as ${\bf x}_s$ 
(${\bf x}_n$). False Positive (False Dismissal) corresponds to a noise (signal) trigger 
wrongly labelled as ${\bf x}_s$ (${\bf x}_n$). 
If $N_{\rm TP}$ and $N_{\rm FP}$ are the numbers of True Positives and False Positives 
evaluated from the output of the classifier at a specific threshold $\hat{p}_t$, then the 
classical Detection Probability (DP) and False Positive Rate (notated as $\alpha$) are 
estimated as
\begin{equation}
DP = \frac{N_{\rm TP}}{N_s},
\end{equation}  
\begin{equation}\label{FPR}
\alpha = \frac{N_{\rm FP}}{N_n}.
\end{equation}
These quantities summarize the performance of the classifier, and can be represented 
visually via the ``receiver operating characteristic'' (ROC) graph.  

The detection probability and false positive rate computed from 
the output of a probabilistic classifier, like the one employed in this project, are a 
function of the score threshold used to label triggers. By varying the threshold 
value from $0$ to $1$, an ROC curve may be created. 
%
Here we use the classifier scores assigned to simulated signals as thresholds for plotting
the ROC.  The false positive rate associated with a signal trigger $\alpha_s$ is 
computed by counting the number of noise triggers whose scores $\hat{p}_n^i$ are 
greater than the score of the signal trigger $\hat{p}_s$:
\begin{equation}\label{FPR_evaluate}
\alpha_s = \frac{1}{N_n}\sum^{N_n}_{i=1} \Theta(\hat{p}^i_n - \hat{p}_s).
\end{equation}

For reasons of computational expense, the amplitude distribution of simulated signals used 
in this study is different from the astrophysical $\rho^{-4}$ distribution expected for 
merging binaries uniformly distributed over space.  
Our set of simulated mergers is, instead, evenly 
distributed in distance from the detector, leading to an expected $\rho^{-2}$ distribution
of signals.
%
Therefore the detection probability for a given threshold $\hat{p}_t$ is \emph{not} 
directly proportional to the number of simulated signals with higher scores; instead, we 
compute a figure of merit proportional to the expected number of detections, $N_d$, by 
performing a weighted sum of signal triggers.  To compensate for the non-astrophysical 
distribution of signal amplitudes we use a weighting inversely proportional to the square of 
the trigger SNR $\rho_s^j$, 
\begin{equation}\label{ROC_weight}
w_{\mathrm{ROC}}^j = \left( \frac{8}{\rho_{s}^j} \right)^2.
\end{equation}
Our estimate of the relative number of detections $N_d$ at a threshold $\hat{p}_s$ is the 
sum of weights $w_{\mathrm{ROC}}^j$ over simulated signal triggers $j$ with scores 
$\hat{p}_s^j$ 
greater than $\hat{p}_s$:
\begin{equation}
N_d(\hat{p}_s) = 
 \sum_{j=1}^{N_s} w_{\mathrm{ROC}}^j \Theta(\hat{p}_s^j - \hat{p}_s). 
\end{equation}
Every point ($\alpha_s$, $N_d$) on the ROC plot then corresponds to classification 
using the score $\hat{p}_s$ of a given signal trigger as a threshold.  
We may also compute the figure of merit for a {\it predefined} set of $\alpha$ values: 
$N_d$ evaluated at a chosen $\alpha$ is the sum of weights $w_{\mathrm{ROC}}^j$ for signal 
triggers whose false positive rates $\alpha_s^j$ evaluated via Eq.~(\ref{FPR_evaluate}) are 
less than or equal to the chosen $\alpha$:
\begin{equation}
N_d(\alpha) = \sum_{j=1}^{N_s} w_{\mathrm{ROC}}^j \Theta(\alpha - \alpha_s^j).
\end{equation}
Note that $N_d$ is only defined up to an arbitrary multiplicative constant; one cannot use
it to predict the \emph{absolute} number or rate of signal detections, only to compare the 
\emph{relative} number of detections between different classifiers, i.e.\ different methods
of assigning scores $\hat{p}$ to triggers ${\bf x}$.

\section{Random Forest as a multivariate classifier}\label{sec:RF}
The Random Forest (RF) is an ensemble classifier consisting of a collection of random 
decision trees~\cite{RandomForest}. 
The RF algorithm employed in this project uses a particular kind of decision tree known as 
a classification tree, which, when trained, predicts the probability of class membership of 
test data points. A classification tree has a structure similar to a flowchart, made up of a 
root node, internal nodes, branches and leaf nodes. Each internal node is connected to a 
parent node and two 
child/daughter nodes via branches, starting with the root node (with no parent) and ending at 
leaf nodes (with no children). During training, each node splits the training data set into 
two sections and sends them along branches to daughter nodes. The branch that a data point 
will follow depends on the splitting condition imposed at the parent node.

In principle, each point in a $p$-dimensional feature space provides $p$ possible splitting
criteria; if this space contains an infinite number of such points, these would correspond 
to an infinite number of ways in which the training data set can be segregated.  In practice, 
not all of those splits will result in unique two-way divisions of the data set. Typically, 
one considers only those $n$ points in feature space that are occupied by the $n$ training 
data points themselves, and chooses the splitting condition from $n\times p$ 
possible attribute values that minimizes the mixing of classes at the daughter nodes. 
We use ``Gini Impurity'' ($I_G$) as a measure of this mixing of classes. The Gini Impurity 
at a node that has data points from $m$ classes is computed using the formula:
\begin{equation}
I_{G} = 1-\displaystyle\sum\limits_{i=1}^m f_{i}^2,
\end{equation}
where $f_{i}$ is the fraction of data points belonging to the $i^{th}$ class. Starting from 
the root node and progressing down the tree node by node, the splitting condition at a node 
is chosen in a way that yields the maximum reduction in Gini Impurity when going from that 
node to its daughter nodes.

A test data point with feature vector ${\bf x}$ pushed into a trained classification tree, 
trickles down node to node based on the splitting conditions imposed at each node, until it 
reaches a leaf node, at which the probability (given ${\bf x}$) of its membership to a class 
$c$, may be estimated. This estimate, which we denote as $\hat{p}(c|{\bf x})$,  
is simply the fraction of training data points at the leaf node 
belonging to class $c$.

A random tree is a straightforward modification to the classification tree. In a random tree, 
the search for the optimum splitting criterion at each internal node occurs over a random 
sub-space of the existing feature space, with only the size of this sub-space being the same 
at each node.  
(Note that the size of the sub-space is a tunable property of the random tree; in fact, 
in our RF implementation we allow the algorithm to search for the 
best split over the entire feature space at each node.)

The random forest algorithm grows multiple random trees, and trains each random tree serially 
on a different ``bootstrapped'' sample of the original training data set - a technique known 
as ``bagging''.  More specifically, given $N$ training samples, each random tree draws a 
bootstrap from this training set by choosing $N$ samples at random with replacement.

A test data point supplied to a trained RF with $T$ random trees is pushed down 
simultaneously into each random tree until it reaches a leaf node. If 
$\hat{p}_t(c|{\bf x})$ is tree $t$'s estimate of the probability that the test data point 
belongs to class $c$, then the prediction by the RF of the same probability is: 
\begin{equation}
\hat{p}_{RF}(c|{\bf x}) = \frac{1}{T}\displaystyle\sum\limits_{t=1}^T \hat{p}_t(c|{\bf x}).
\end{equation}

We choose the random forest method for this classification problem as it is straightforward
to implement, computationally manageable, does not require special transformations of the 
input data and yields results that are relatively insensitive to choice of hyperparameters, 
e.g.\ number of trees, splitting criterion, leaf size (see Section~\ref{sec:results} for more 
details of these choices).  We have also investigated other classification algorithms such as
Nearest Neighbours and Support Vector Machine, which yield comparable results but are less
robust to parameter changes.

\section{Feature generation and selection for an inspiral search}\label{sec:features}

\subsection{Inspiral search triggers}
%
\begin{figure*}
\centering
\subfigure[]{
  \includegraphics[width=.45\linewidth]{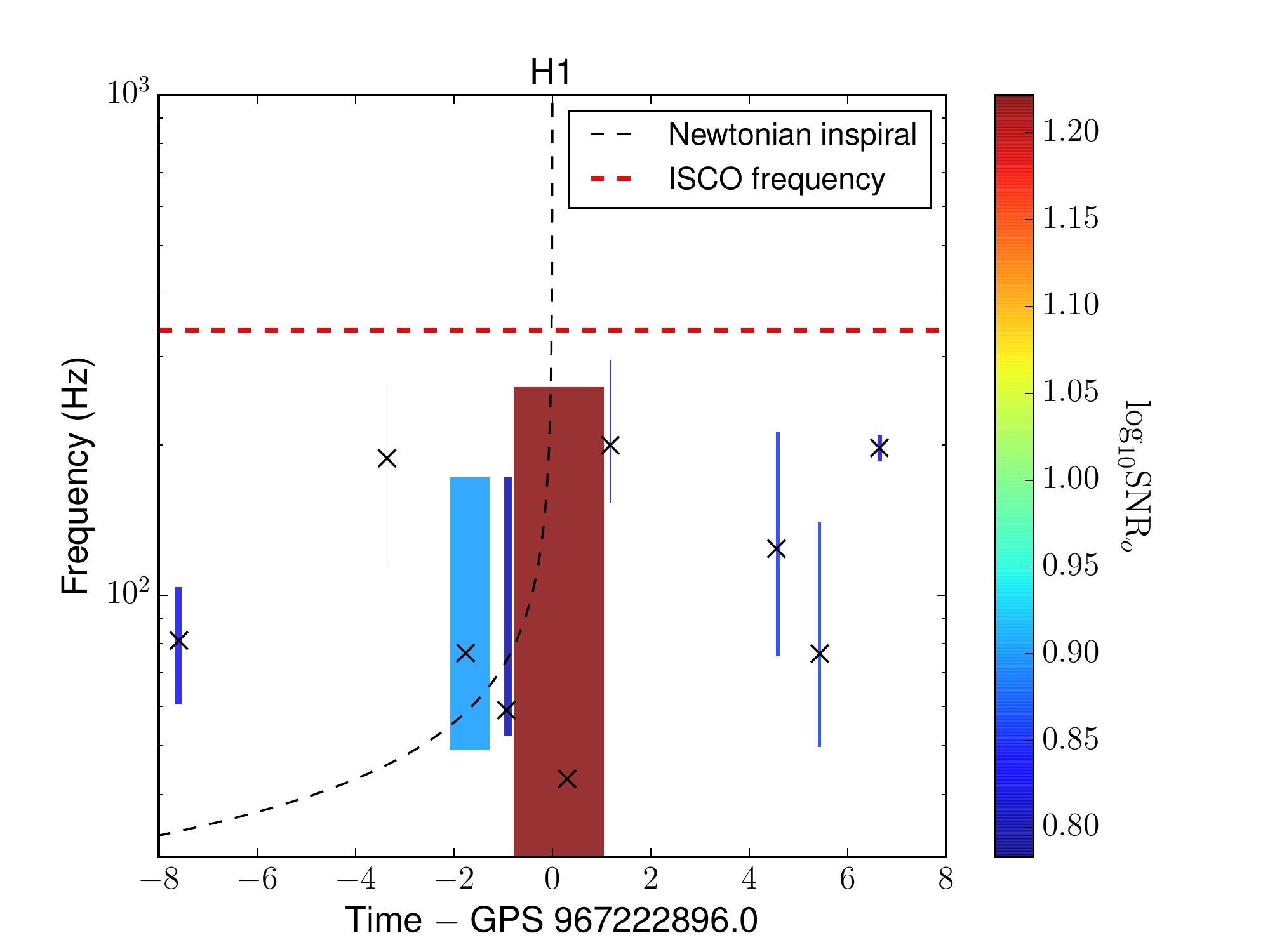}
  \label{OmicronNoise}}
\quad
\subfigure[]{
  \includegraphics[width=.45\linewidth]{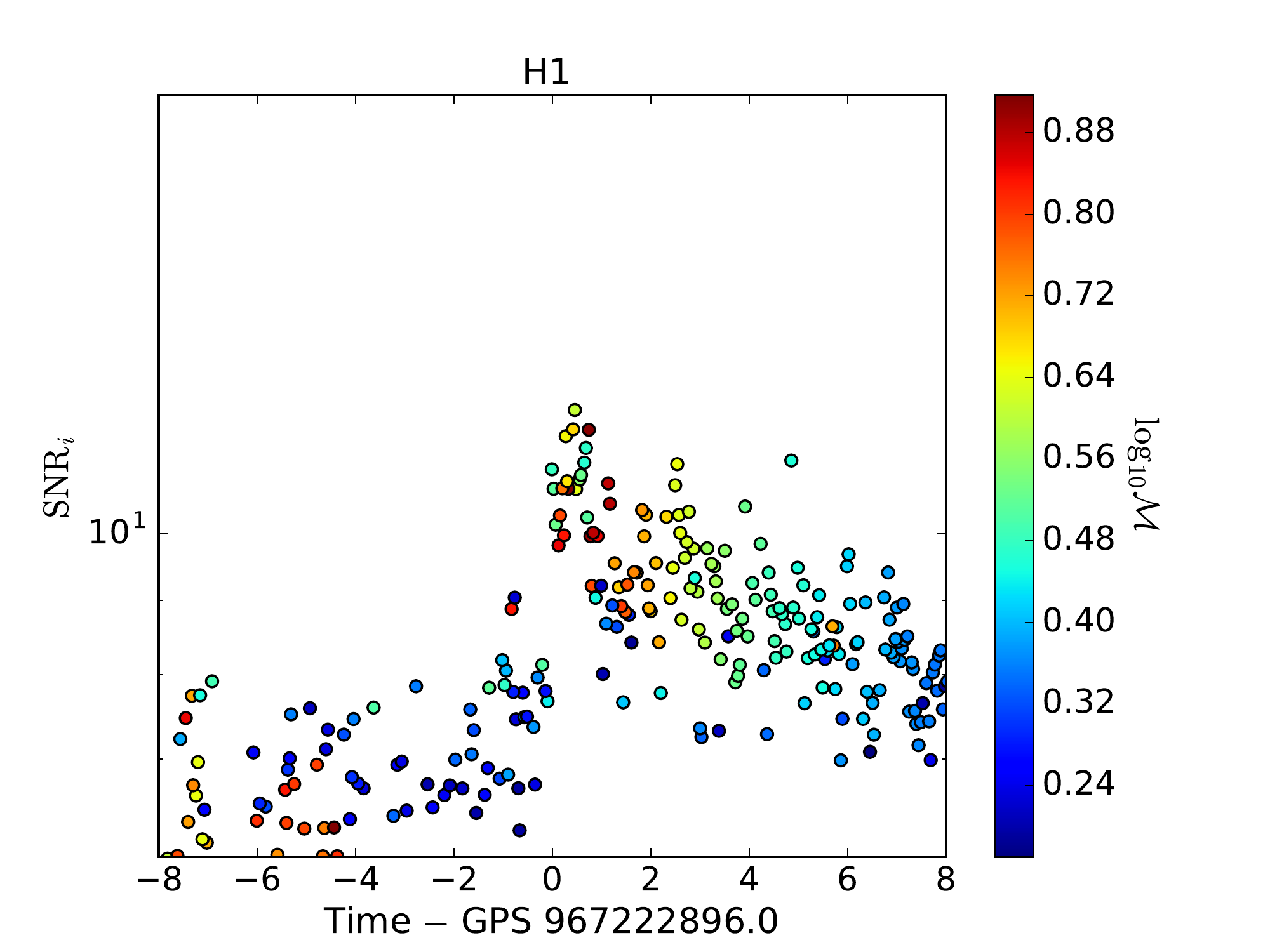}
  }
\caption{Omicron (left) and inspiral (right) clustered triggers within a 16-second time window 
centred on a high-SNR inspiral search trigger due to a noise artefact in LHO recolored mock data. 
The rectangles in the omicron plot indicate the time and frequency limits of trigger (tile)
clusters, while the $\times$ sign locates the highest-power tile in each cluster.  In the 
inspiral trigger plot the color indicates chirp mass $\mathcal{M}$. 
Note that the loudest omicron triggers, while overlapping the 
track of the high-SNR inspiral search trigger, also cover much wider regions of time-frequency 
space and are not peaked on the inspiral track.  The inspiral triggers show elevated SNR over 
several seconds and over a wide range of $\mathcal{M}$.
\label{NoiseFollowup}
} 
%
\subfigure[]{
  \includegraphics[width=.45\linewidth]{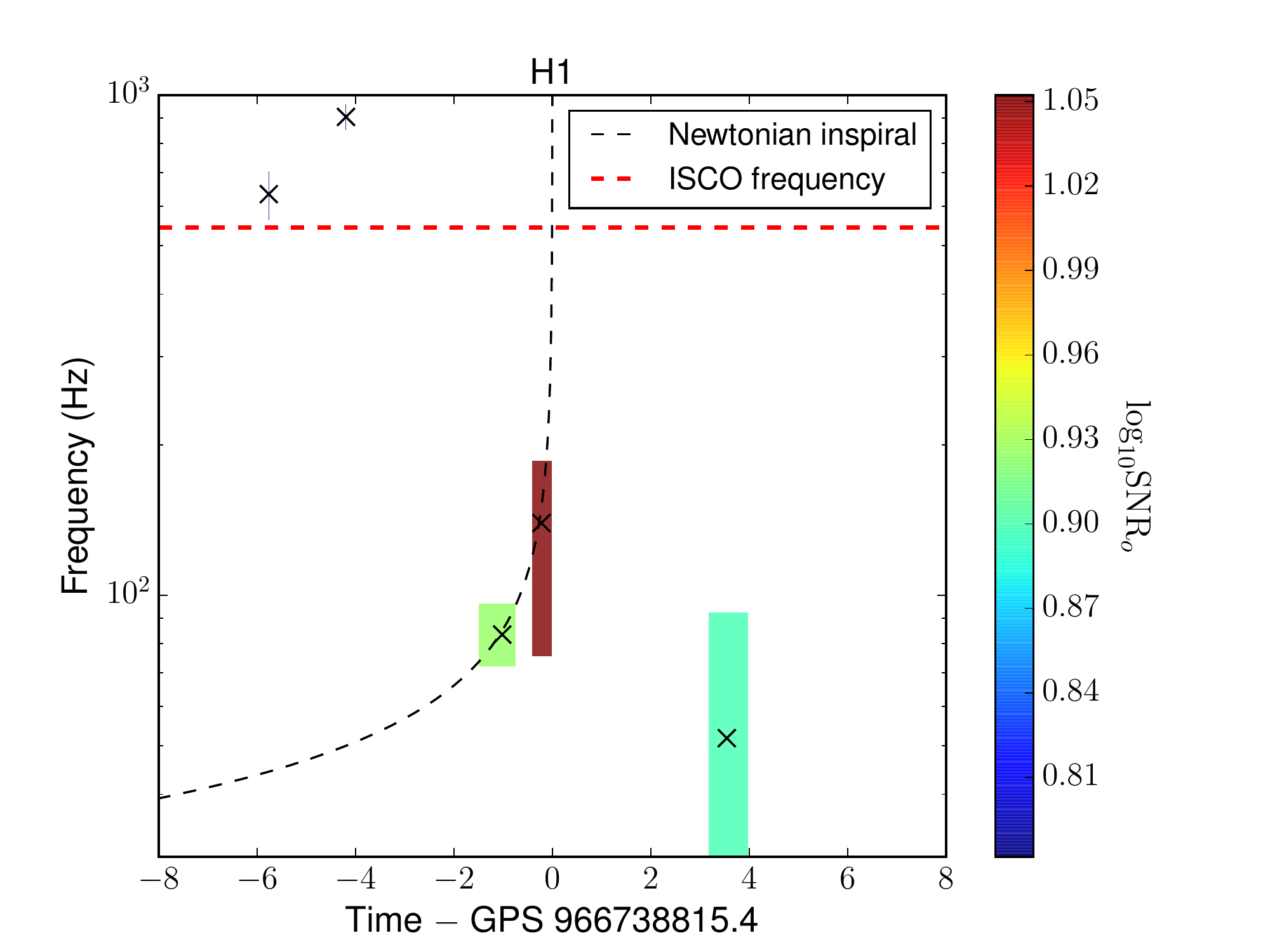}
  \label{OmicronSignal}}
\quad
\subfigure[]{
  \includegraphics[width=.45\linewidth]{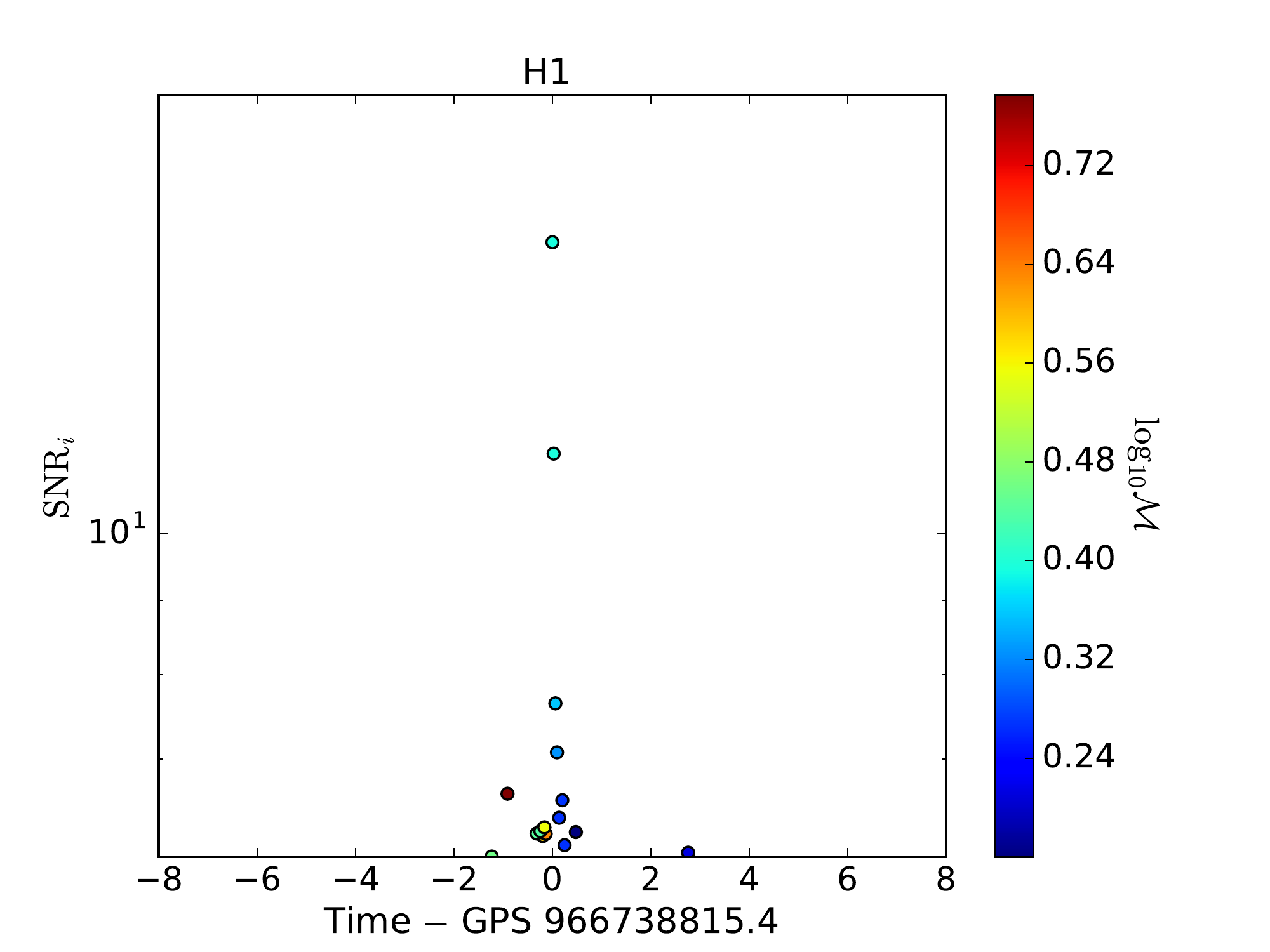}
  }
\caption{Omicron (left) and inspiral (right) clustered triggers within a 16-second time window 
centred on a simulated signal added to LHO recolored mock data.  The loudest omicron triggers lie 
around the signal's time-frequency track and are peaked \emph{on} the track.  The omicron trigger
between approximately $+3$ and $+4$\,s is due to an unrelated lower-frequency glitch which does
not generate an inspiral trigger.  The inspiral triggers have low SNR except very 
close to the signal coalescence time and within a narrow $\mathcal{M}$ range.} 
\label{SignalFollowup}
\end{figure*}
The data used for training and evaluation of the classifier were generated principally by 
applying a templated matched filter search~\cite{FindChirp} to realistic simulated early 
Advanced LIGO data covering approximately $10^6\,$s for each of two observatories, Hanford 
(LHO) and Livingston (LHO).\footnote{We will refer to this data, generated by re-coloring LIGO 
S6 data~\cite{S6DC} to have an average power spectral density matching a projected early 
Advanced LIGO sensitivity~\cite{LIGOT1200307}, as 'recolored mock data'.}  The bank of 
templates used for filtering and the implementation are described in~\cite{Canton:2014ena}, 
where our analysis uses the `non-spinning search' configuration.  The template waveforms 
are restricted TaylorF2 approximants for binaries with non-spinning components of BH mass 
($m_1$) ranging over $3$--$15\,M_\odot$ and second component (either a NS or BH) ranging 
from $1\,M_\odot$ up to equal mass $m_2\leq m_1$; total binary mass is restricted to 
$M\leq18\,M_\odot$, resulting in $\sim28\,000$ templates.  Although the target space of 
signals for this bank comprises NS-BH systems with a NS mass up to $3\,M_\odot$, we allow 
spinning signals to be recovered by non-spinning templates with similar chirp mass but 
closer to the equal-mass boundary; i.e.\ we tolerate a bias in the recovered mass ratio. 

Triggers are generated by finding local maximum of matched filter SNR $\rho(t)$ in each 
template, where $t$ denotes the coalescence time of the inspiralling system, above a 
predetermined threshold; here we choose $\rho>\rho_{\rm th}=6$.  To 
reduce the incidence of highly-correlated triggers from templates with large overlaps, 
clustering over the template bank was performed: triggers for which there exists a
higher-SNR trigger in another template within a time window of 25\,ms are discarded, 
keeping only those with the highest SNR \emph{over all templates} inside the window. 
A $\chi^2$ signal consistency test was also calculated for the triggers~\cite{Brucechisq}
with the standard choice of $p=16$ frequency bins, although its value was not 
used for our principal multivariate classifier results.  Finally, triggers falling within 
times marked as affected by instrumental and environmental disturbances with a known 
coupling to the strain channel (``Category 2 veto''~\cite{S5DQ}) and times of hardware 
injections, i.e.\ signals simulated by actuating the interferometer optics, are removed 
from the data set. 

The recolored mock data provide a set of noise events, each specified by a GPS time, 
template binary component masses, and matched filter SNR value (and, where calculated, 
a $\chi^2$ value); the $10^6\,$s of recolored mock data provide $\sim480\,000$ noise events
from LHO and $\sim320\,000$ from LLO.  We also require corresponding events that model 
astrophysical NS-BH signals in order to train the classifier and evaluate its sensitivity.  
We add simulated NS-BH signals to the recolored mock data within the search pipeline 
(`software injections') using the {\tt SpinTaylorT2} approximant implemented in {\tt 
LALSimulation}~\cite{LALSuite}.  We inject a population covering the NS-BH space evenly with 
component masses distributed uniformly on $2\leq m_1/M_\odot \leq16$ and $1\leq m_2/M_\odot 
\leq 3$; selecting only events associated with the simulated signals we obtain $\sim6100$ 
events from each observatory's data. 

Dimensionless spin magnitudes were 
distributed uniformly on $(0,1)$ for the BH component and $(0,0.05)$ for the NS, while spin 
directions were uniform on the sphere.  Thus, many simulated signals exhibited orbital 
precession, which is expected to affect the recovery of their SNR~\cite{Harry:2013tca} in 
non-spinning templates, as well as in the spin-aligned (non-precessing) 
templates~\cite{Ajith:2012mn,Privitera:2013xza,Canton:2014ena} used in recent searches of 
Advanced LIGO data~\cite{CBCCompanion}.  Mismatch between precessing signals and non-precessing 
templates will also lead to larger chi-squared values, which will reduce the effectiveness of 
the chi-squared test in distinguishing signals from glitches. 

Note that there is significant theoretical uncertainty in NS-BH signal 
waveforms~\cite{Nitz:2013mxa} due both to the incompleteness of the post-Newtonian expansion 
and the technical difficulty of numerically simulating unequal-mass systems evolving well 
before merger.  Thus, there may be some (moderate) mismatch between \emph{real} NS-BH signals 
and our search templates (whether non-spinning or spin-aligned): when evaluating detection 
methods it may be desirable to have some degree of mismatch between \emph{simulated} signals and 
search templates. 

\subsection{Omicron: a sine-Gaussian glitch characterization algorithm}

While compact binary inspiral signals can be approximately described by post-Newtonian 
theory, other types of transient GW signal, described in general as ``bursts'', 
are in general more computationally expensive to simulate and have more systematic 
uncertainty due to complicated processes in the physics of super-dense matter.  
Thus, many searches for bursts in GW data have used methods which rather than using
signal templates, instead aim to detect generic 
(weakly-modelled) transient excess power events in the strain time series from one
or more GW detectors~\cite{Klimenko:2015ypf,Kanner:2015xua,Lynch:2015yin}.  

Omicron~\cite{OmicronRobinet} is a single-detector burst trigger generator employing 
the Q-transform~\cite{ShourovThesis} which uses windowed sinusoids that form an 
over-complete set of basis functions of varying durations, covering the time-frequency 
plane within a specified frequency range. 
Each single-detector data stream is whitened using a PSD produced via the mean-median
method, and is then projected onto the basis functions; any projection with energy over
a pre-determined threshold results in the production of an Omicron trigger.  The basic
event production 
is thus similar to a template matched filter~\cite{ShourovThesis}. 

It has been shown through approximations and numerical computations that the
inner product of inspiral and sine-Gaussian waveforms can be large when the
parameters of the two signals satisfy a specific relation~\cite{DalCanton:2013joa}.  
Intuitively, the inner product is large when the
time-frequency supports of the signals overlap significantly, in particular
when the time-frequency curve traced by the inspiral passes through the point
$(\tau,\phi)$, $\tau$ and $\phi$ being the center time and frequency of the
sine-Gaussian. On the other hand, the signals are almost orthogonal when their
time-frequency supports are very far apart. Thus, when Omicron is used in data
containing a loud inspiral signal, we expect a sequence of sine-Gaussian triggers
``covering'' the time-frequency curve associated with the inspiral. Conversely,
when an inspiral template bank is used against data affected by a sine-Gaussian
glitch, a sequence of triggers is produced by templates whose time-frequency
curve overlaps with the sine-Gaussian.

Omicron uses windowed sinusoids uniquely defined by their center frequency $\phi$, center time 
$\tau$, and quality factor $Q$, which may be viewed as ``tiles'' of constant $Q$ in the 
time-frequency plane. The bandwidth of the tile is defined as $2\sqrt{\pi} \phi/Q$, while the 
duration of the tile is the inverse of the bandwidth due to the uncertainty relation. 
Gaussian windowed sinusoids provide the maximum possible resolution for the matched filter 
allowed by the time-frequency uncertainty relation, and are therefore the basis 
functions of choice in principle~\cite{ShourovThesis}. However, the infinite extent of Gaussian 
sinusoids is incompatible with the periodic window sequence required when performing a discrete 
Q transform. Therefore, in practice, the near-minimum uncertainty bi-square window of finite 
extent is used~\cite{ShourovThesis,S5BurstPaper}.

The set of basis functions may be viewed as lattice points in signal space with axes 
as frequency, time and quality factor. The distance between lattice points is 
judiciously chosen so as to ensure that the mismatch between an arbitrary burst within the space 
spanned by the basis functions, and the windowed sinusoids, does not exceed a predefined threshold. 
We here briefly review the generation of Omicron triggers to fix notation for their use in the 
classifier.

For each lattice point, the projection of the data stream $x(t)$ onto the Omicron basis function 
produces a Q-transform coefficient $X$:
\begin{equation}
X(\tau,\phi,Q) = \int_{-\infty}^{+\infty}x(t)W(t-\tau,\phi,Q)e^{-i2\pi\phi t}dt ,
\end{equation}
where $W(t-\tau,\phi,Q)$ is the bi-square window function. The tile energy is measured by the 
squared magnitude of the Q transform coefficients, $|X|^2$. 
In the absence of localized excess energy, the expectation of the mean tile energy can be 
shown to be 
\begin{equation}
\langle |X_n(\tau,\phi,Q)|^2 \rangle = \frac{1}{2} \int_{0}^{+\infty} S_n(f) |W(\phi-f)|^2 df,
\end{equation}
where $S_n(f)$ is the one-sided PSD. 
In Omicron, the mean tile energy is computed for each Q plane and for each frequency bin. An outlier 
rejection technique~\cite{ShourovThesis} is used to exclude localized and loud bursts in the data. 
Moreover, the frequency bin is narrow enough so the PSD is assumed to be constant over the bin, in 
which case $\langle |X_n(\tau,\phi,Q)|^2 \rangle \simeq 1/2 S_n(\phi)$.

For each tile, Omicron computes the Q-transform coefficient, $|X(\tau,\phi,Q)|^2$, from which the
SNR is derived as
\begin{equation}
\mathrm{SNR_{omi}^2}(\tau,\phi,Q) = 
 \frac{|X_e(\tau,\phi,Q)|^2}{\langle |X_n(\tau,\phi,Q)|^2 \rangle} -1 \simeq 
 \frac{2|X_e(\tau,\phi,Q)|^2}{S_n(\phi)} -1.
\end{equation}
An Omicron trigger is defined as a tile with a SNR above a given threshold.  

To reduce the vast amounts of triggers produced during an Omicron matched filtered search, we 
employ a technique known as ``clustering'' which lumps together tiles produced with small time 
separations.
Two 
tiles above threshold are 
clustered together if the difference between the end time of the first and the start time of the 
second is smaller than $100\,ms$; clustering is continued until no more tiles can be added.  
The resulting cluster is parameterized by the peak time 
($t_{\rm peak}$), peak frequency ($f_{\rm peak}$) and peak SNR ($\mathrm{SNR}_{\rm omi}$)
of the highest-SNR tile in the cluster.  The start (end) time of the cluster is simply the 
start (end) time of the tile with the smallest (largest) start (end) time value, and is denoted by 
$t_s$ ($t_e$). The start (end) frequency, $f_s$ ($f_e$), is similarly defined. The cluster 
bandwidth $\sigma_f$ and duration $\sigma_t$ may then be trivially computed.


These output values may be visually represented by a 
two-dimensional 
time-frequency plot where each clustered Omicron trigger is plotted as a rectangle of height equal 
to its bandwidth $\sigma_f$ and width equal to the duration $\sigma_t$.  The color of the rectangle 
indicates the trigger's peak SNR $\mathrm{SNR_{omi}}$ and the $\times$ symbol within each rectangle 
indicates the peak time and frequency of the Omicron trigger. 

Figures~\ref{NoiseFollowup} and~\ref{SignalFollowup} show two such plots, one centred on the time 
of an inspiral noise trigger, the other centred on the time of a simulated inspiral signal trigger. 
In addition to the 
Omicron triggers, each of these plots show a Newtonian-order time-frequency trajectory for the 
leading order GW emission of a binary system with mass parameters given by the template where the 
respective inspiral trigger was seen.

The Omicron triggers of figure \ref{OmicronNoise} are all short-lived ($\sigma_t \lesssim 1 sec$) 
and lie scattered around the inspiral track, except for the loudest glitch which appears close 
to the inspiral trigger time and partially overlaps with the track. However, its central frequency 
is noticeably away from the inspiral track, an indication that it is likely not an inspiral signal 
trigger.

When the Omicron algorithm encounters data containing a (simulated) inspiral signal, the inspiral
track corresponding to this signal gets covered by a series of Omicron triggers.  We see two such
triggers in figure \ref{OmicronSignal}. 
Not unexpectedly, the loudest of these triggers sits on the inspiral track with a peak time close 
to the time of the inspiral signal trigger, i.e.\ slightly before the time of coalescence (as seen
at the detector).

\subsection{Classification features}

Using the properties of the inspiral and Omicron triggers, we pass to the random forest
algorithm a series of seven inspiral features and fourteen Omicron features, on which 
the RF will train itself and use to classify noise and signal triggers.

The following lists the inspiral features, briefly explaining each one:
\begin{itemize}
\item $\mathrm{SNR_{insp}}$: The inspiral matched filter SNR of a trigger. 
\item $\mathcal{M}$: The chirp mass of the template used to identify the candidate event, 
given by Eq.~(\ref{eq:mchirp})
\end{itemize}

For the following features, we consider 0.2 second, 2-second, and 20-second time windows 
$\delta t_{0.1}, \delta t_{1}, \delta t_{10}$ centred on the time at which the inspiral candidate 
event occurred, and determine the maximum signal to noise ratio values $\mathrm{SNR_{insp}^{0.1}}, 
\mathrm{SNR_{insp}^{1}}, \mathrm{SNR_{insp}^{10}}$ from the SNRs of the triggers within each of the 
time windows.  
\begin{itemize}
\item $n_{1, 10}$: The number of inspiral triggers within $\delta t_{1, 10}$
\item $\mathrm{\Delta SNR_{insp}^{0.1, 1, 10}}$: The difference between the maximum SNR value in 
each of the time windows $\delta t_{0.1,1,10}$ and the SNR of the trigger on which the window is 
centred:
\begin{equation}
\mathrm{\Delta SNR_{insp}^{0.1, 1, 10}} = \mathrm{SNR_{insp}^{0.1, 1, 10}}-\mathrm{SNR_{insp}}
\end{equation}
\end{itemize}

The following set of fourteen Omicron features rely entirely on the time at which the inspiral 
trigger was found. The properties of the loudest Omicron trigger within a time window of $20$\,s 
duration centred on the inspiral trigger time are used as the first five Omicron features:
\begin{itemize}
\item $\mathrm{SNR_{omi}^{10}}, f^{10}_{s}, f^{10}_{e}, \tau^{10}, \phi^{10}$: The SNR, start 
frequency, end frequency, center time, center frequency of the loudest Omicron trigger within 
the $20$\,s time window.
\end{itemize}
\begin{figure}[htb]
\centering
\includegraphics[trim=30 20 20 40, clip, width=0.47\textwidth]{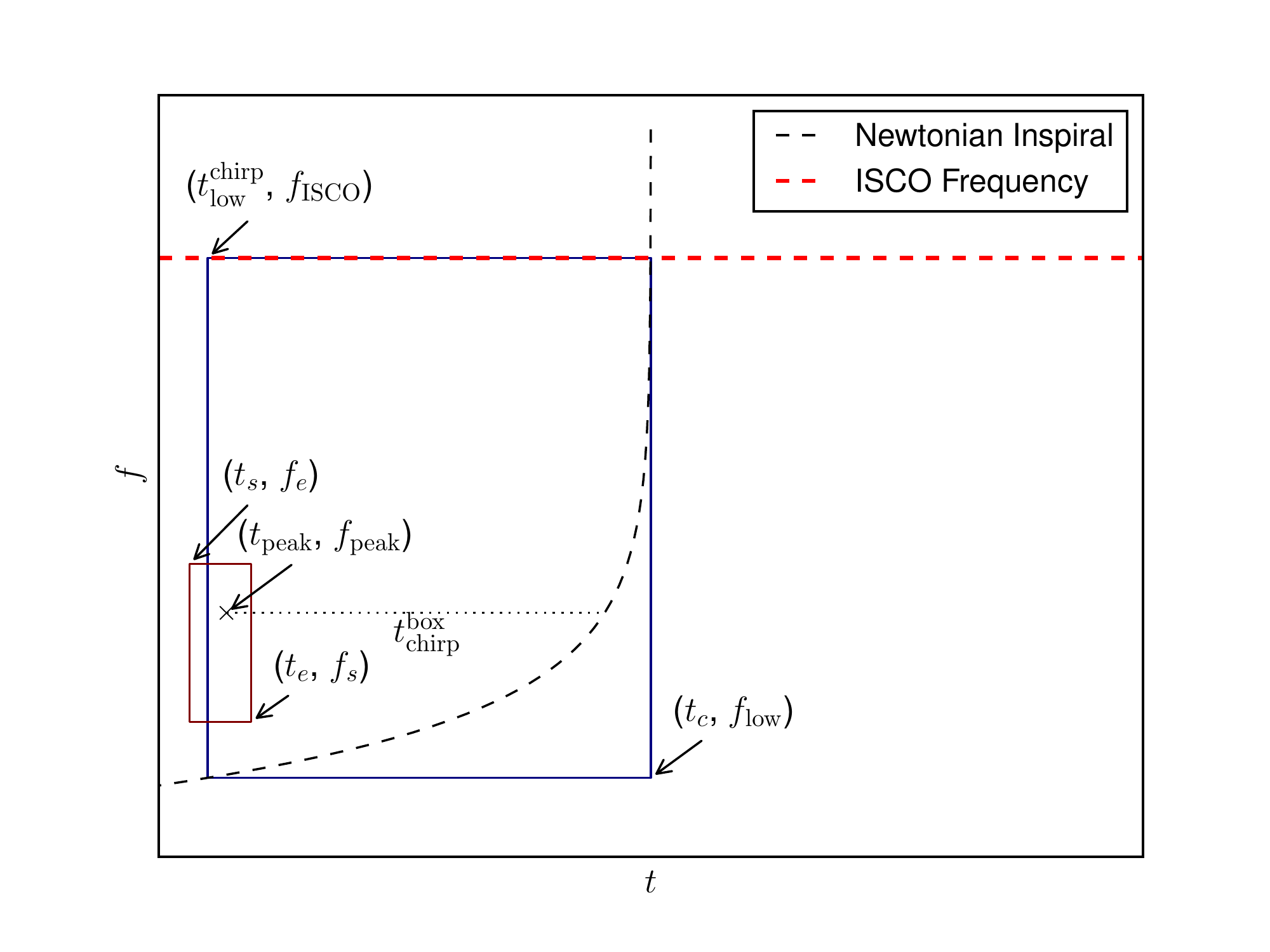}
\caption{The time-frequency box associated to an inspiral search trigger (candidate signal) 
with coalescence time $t_c$ 
and highest GW frequency $f_{\rm ISCO}$.  
The inspiral trigger is a maximum of the matched filter time series, where the filter has 
support between frequencies $f_{\rm low}$ and $f_{\rm ISCO}$.}
\label{fig:feature_box}
\end{figure}

The remaining nine Omicron features are constructed using a time-frequency ``box'' and an 
inspiral ``track'' drawn in the time-frequency plane by assuming $t_c$ to be the coalescence 
time of a binary with chirp mass $\mathcal{M}$ given by the inspiral template: see 
Figure~\ref{fig:feature_box}.  The lower right corner of the box 
is the point $(t_c, f_{\mathrm{low}})$, $t_c$ being the time of the inspiral trigger 
and $f_{\mathrm{low}} = 30Hz$, which is close to the lower end of LIGO's frequency range of 
detectability.  The intersection between the inspiral track and the line of constant frequency 
$f = f_{\mathrm{low}}$ gives the lower left corner of the box 
$(t_{\mathrm{low}}^{\mathrm{chirp}}, f_{\mathrm{low}})$. 
The upper right corner of the box is $(t_c, f_{\mathrm{ISCO}})$, $f_{\mathrm{ISCO}}$ being the 
GW frequency corresponding to the innermost stable circular orbit, computed using the inspiral 
template's chirp mass $\mathcal{M}$ and symmetric mass ratio $\eta$. The upper left corner is 
the intersection of the line of constant frequency $f = f_{\mathrm{ISCO}}$ and $t = 
t_{\mathrm{low}}^{\mathrm{chirp}}$.

The set of Omicron triggers that lie within the box are first identified.  The properties of 
these triggers have to satisfy the conditions $\tau > t_{\mathrm{low}}^{\mathrm{chirp}}, 
t_s < t_c, f_{e} > f_{\mathrm{low}}, f_{s} < f_{\mathrm{ISCO}}$. This gives us the sixth 
Omicron feature:
\begin{itemize}
\item $n_{\mathrm{box}}$: The number of Omicron triggers found within the time-frequency box. 
\end{itemize}

We then determine the loudest Omicron trigger amongst the $n_{\mathrm{box}}$ triggers within 
the box, and construct the remaining Omicron features from the properties of this loudest 
trigger:
\begin{itemize}
 \item $\mathrm{SNR_{omi}^{box}}, t_s^{\mathrm{box}}, t_e^{\mathrm{box}},
\sigma_t^{{\mathrm{box}}}, f_s^{\mathrm{box}}, f_e^{\mathrm{box}}, \phi^{\mathrm{box}}$: 
The SNR, start time, end time, duration, start frequency, end frequency, center frequency of 
the loudest Omicron trigger in the box.
\item $t_{\mathrm{chirp}}^{\mathrm{box}}$: Let $t_{\mathrm{th}}$ be the delay between the 
coalescence time $t_c$ and time-coordinate of the intersection point between the inspiral track 
and the line $f = \phi^{\mathrm{box}}$.  We then define
\begin{equation}
t_{\mathrm{chirp}}^{\mathrm{box}} = t_{\mathrm{th}}-(t_c- \tau^{\mathrm{box}}),
\end{equation} 
where $\tau^{\mathrm{box}}$ is the center frequency of the loudest Omicron trigger in the box. 
$t_{\mathrm{chirp}}^{\mathrm{box}}$ is thus a measure of the temporal separation between the 
inspiral track and loudest Omicron trigger.
\end{itemize}

\section{Random forest tuning and evaluation}\label{sec:results}
\noindent

In this study we employ the random forest (RF) machine learning algorithm to classify triggers 
derived from the inspiral matched filter search.  We consider each trigger as belonging either 
to the class of ``noise'' or ``signal''; the trained RF outputs an estimate of the probability 
of class membership for each trigger supplied to it. 
The RF has a number of adjustable parameters which may affect the accuracy of this estimate 
and the computational cost of training and classification.  Here we briefly summarize the 
parameters that were used in our runs; these were arrived at after stages of grid search on a 
smaller data set.  We give both the names used internally in the Python 
machine learning package \texttt{scikit-learn}~\cite{scikit-learn} 
which we use in this study, and a condensed notation used for labelling our plots.
\begin{itemize}
\item \texttt{n\_estimators} ($N_t$): The number of random trees that the RF algorithm is 
asked to produce. While increasing the number of trees increases the accuracy of the RF's 
output, it also enlarges computational costs. Gain in accuracy by adding more trees 
ultimately starts to saturate, as a consequence of the fact that the training data set is 
of finite size; after a point, additional trees cannot extract distinct information from 
the training data relative to their predecessors (see e.g.~\cite{Baker:2014eba}).  We do 
not see significant improvements from values above $N_t = 128$, which is used for all RF 
implementations here. 
\item \texttt{criterion}: There are multiple ways by which the quality of a split at a node in a 
tree may be measured. The ``worse'' the quality of a split, the larger the mixing of classes of
data points at daughter nodes. We use Gini impurity to measure this mixing of classes 
(\texttt{scikit-learn} also allows the choice of using Shannon entropy). 
\item \texttt{min\_samples\_split}: The minimum number of data points 
required to split an internal node into two daughter nodes.  We use the default value $2$. 
\item \texttt{min\_samples\_leaf} ($l_{\rm min}$): The minimum number of data points that a 
daughter node must have. Any split that produces a daughter node with fewer than $l_{\rm min}$ 
data points is rejected. Reducing $l_{\rm min}$ increases the accuracy of the RF classifier, 
although a too small value of $l_{\rm min}$ may make the RF prone to over-fitting.  Hence this
is the main parameter we use to regulate tree size and complexity. 
\item \texttt{samples\_weight}: As discussed in Section \ref{sec:GeneralFramework}, it is 
worthwhile to assign weights to signal triggers when training the RF, to model a more  
astrophysically realistic distribution. To that end, the noise triggers are all set to be 
equally weighted (unit weights) and the signal triggers are weighted according to the formula
\begin{equation}
w_s = \left(\frac{w}{\rho}\right)^2,
\end{equation}
where $w$ is a tunable parameter; thus, the smaller the signal SNR, the larger the weight.
We also evaluate the RF with the signal triggers all equally weighted; this case will be 
described as 
``unit weights''.
\end{itemize}

Proceeding as described in section \ref{sec:GeneralFramework}, we generate ROC plots 
using the RF's output evaluated on a set of noise and signal triggers, using two-fold 
cross validation, and compute the $N_d$ for a predefined set of $\alpha$s. 
We repeat the evaluation five times, using the same set of predefined $\alpha$s but a different 
realization of the two-fold splitting.  For each $\alpha$, we 
then plot the mean $N_d$ value and the standard deviation, which we use as a proxy for the 
statistical error at each point on the ROC. 

%
\begin{figure*}[ht]
\subfigure[\ LHO data, all triggers weighted equally]{%
\includegraphics[width=.44\linewidth]{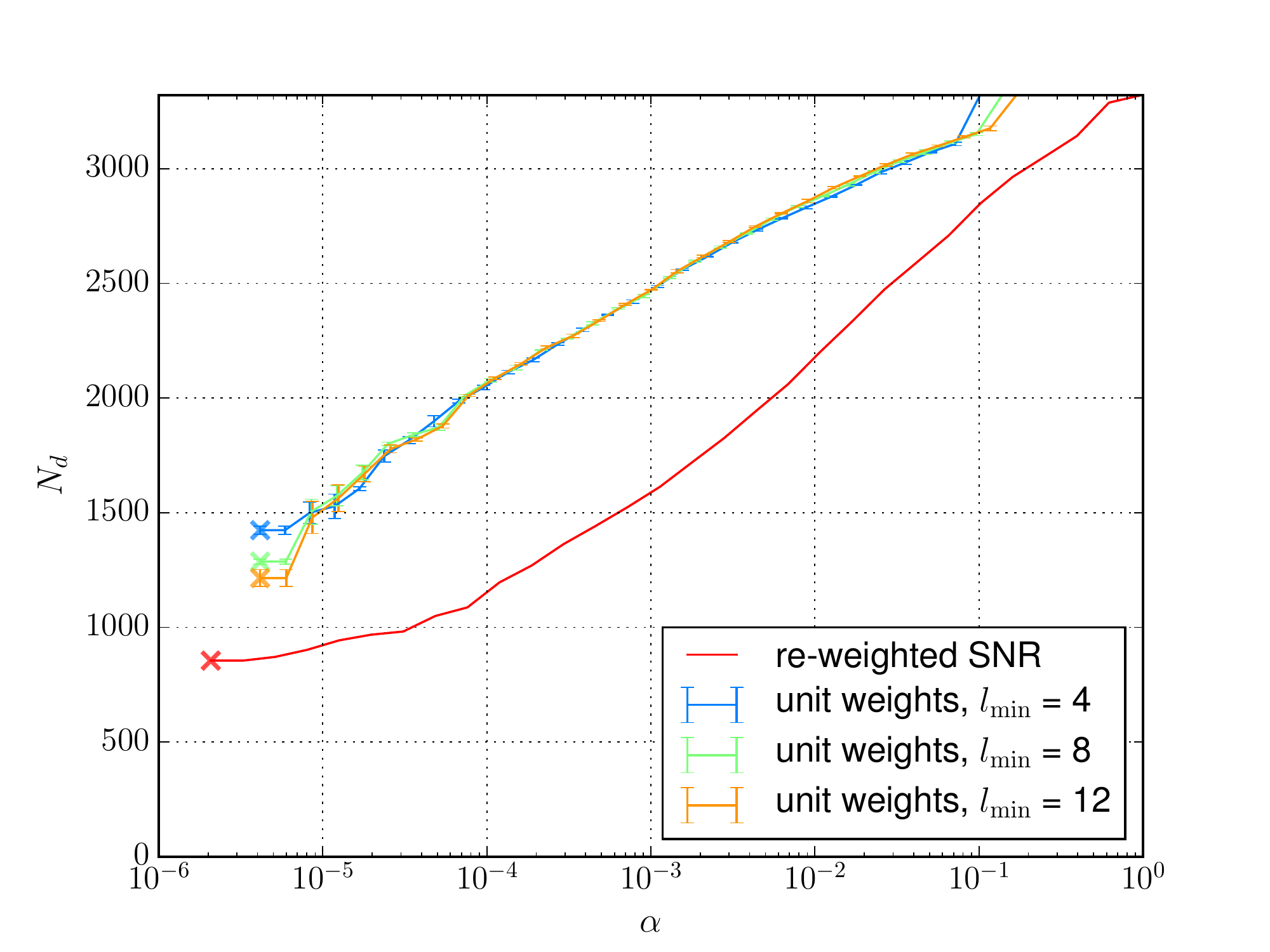}
\label{fig:H1_ROC_w1}}
%
%
\subfigure[\ LHO data, signals weighted as $w_s=10/\rho^2$]{%
\includegraphics[width=.44\linewidth]{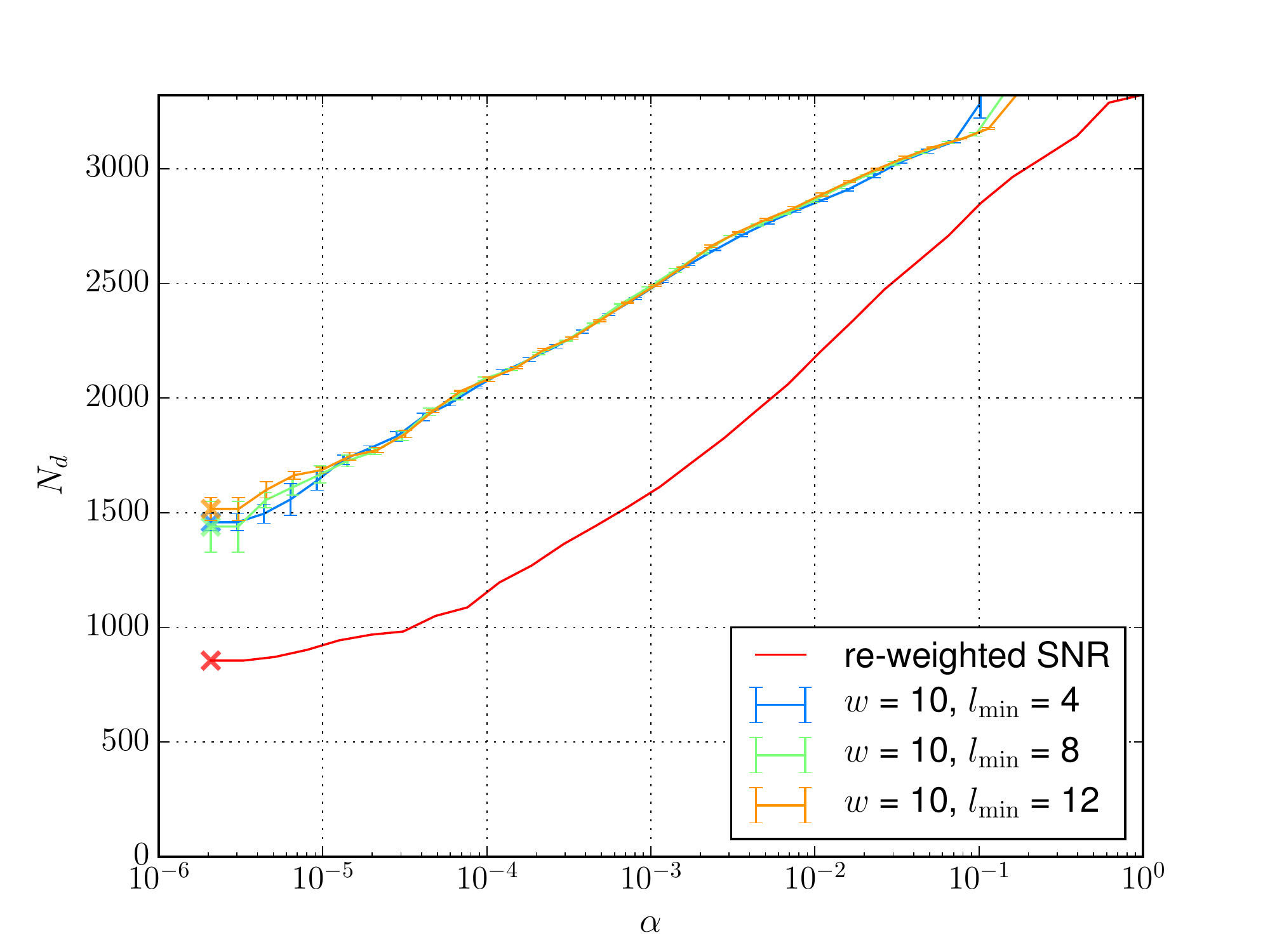}
\label{fig:H1_ROC_w10}}
\subfigure[\ LLO data, all triggers weighted equally]{%
\includegraphics[width=.44\linewidth]{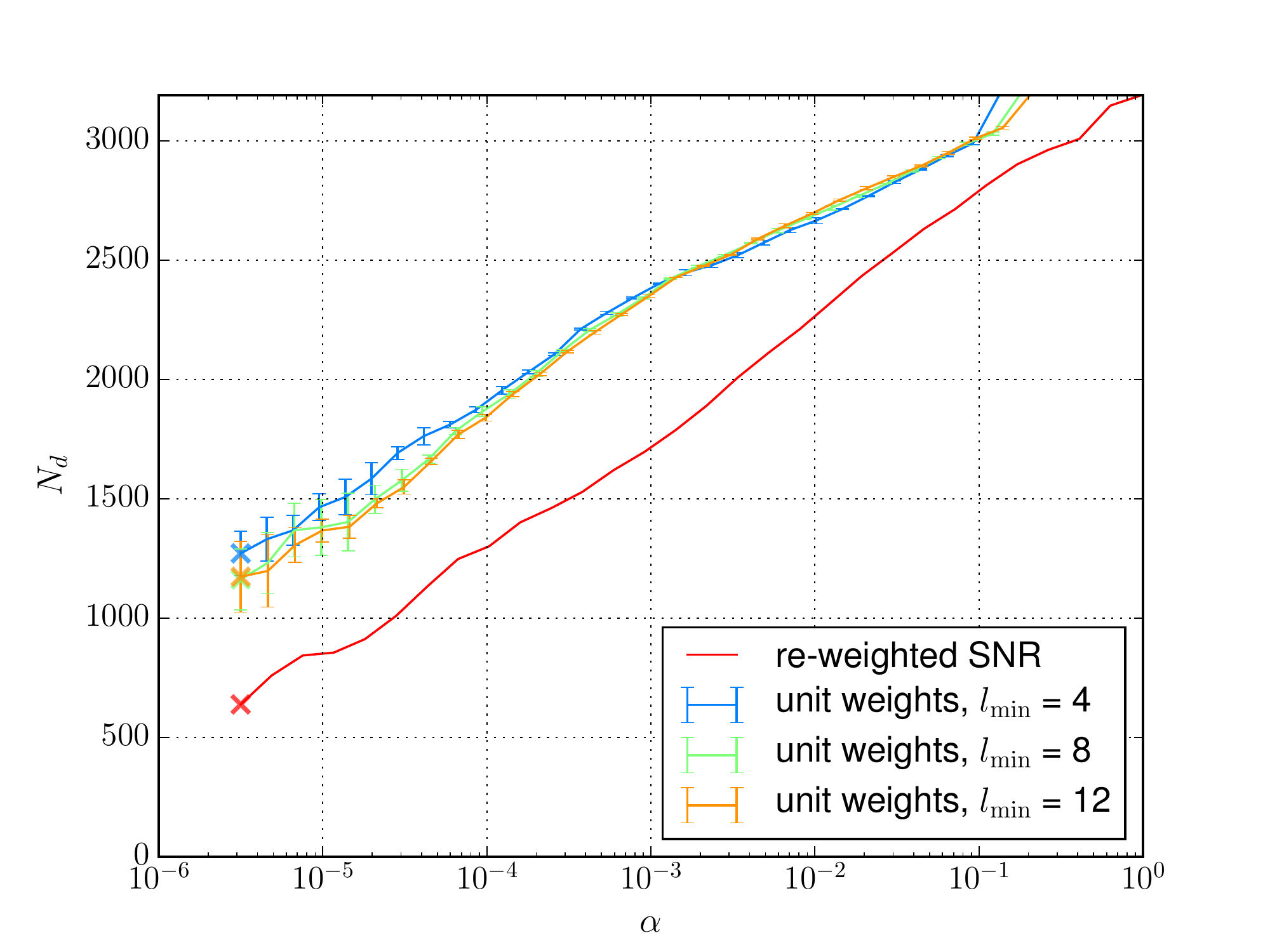}
\label{fig:L1_ROC_w1}}
%
%
\subfigure[\ LLO data, signals weighted as $w_s=10/\rho^2$]{%
\includegraphics[width=.44\linewidth]{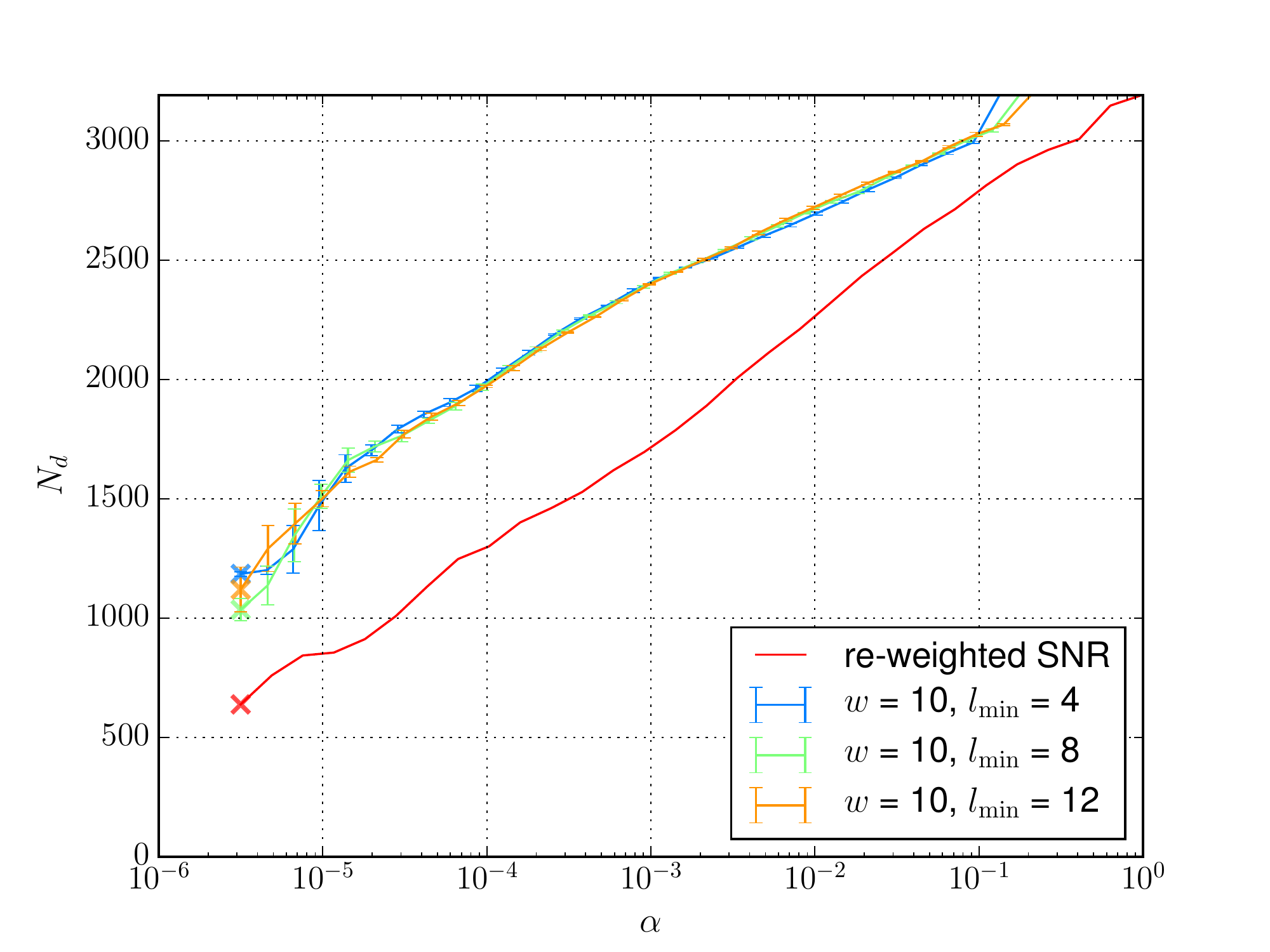}
\label{fig:L1_ROC_w10}}
\caption{ROC plots for the random forest (RF) classifier for different forest parameters. 
Each panel shows the expected number of detected signals $N_d$ (in arbitrary units) vs.\ 
the type I error rate (false positive rate) $\alpha$, estimated by 2-fold cross-validation on 
trigger sets derived from realistic early Advanced LIGO noise with simulated spinning 
compact binary inspiral signals.  For each plot, the red line shows the ROC of the standard 
matched filter re-weighted SNR, while the colored lines with error bars show ROCs for the
RF classifier with different minimum leaf sizes.  Each column corresponds to a different 
choice of relative weights for signal and noise triggers in training the classifier.  All 
RFs used $N_t = 128$ trees.}
\label{fig:H1_L1_ROC_plots}
\end{figure*}

Fig.~\ref{fig:H1_L1_ROC_plots} shows that the RF classification yields a higher expected 
detection rate than the re-weighted SNR statistic by up to a factor of two at low false 
positive rates $\alpha \sim 10^{-5}$.
The statistical errors in the RF curves estimated from different random cross-validation 
realizations are small.
The efficiency of the RF classification for a given set of data also does not change greatly 
upon changing the RF parameters, which is an indication of the robustness of the RF algorithm.  

\begin{figure}[htb]
\centering
\includegraphics[width=0.5\textwidth]{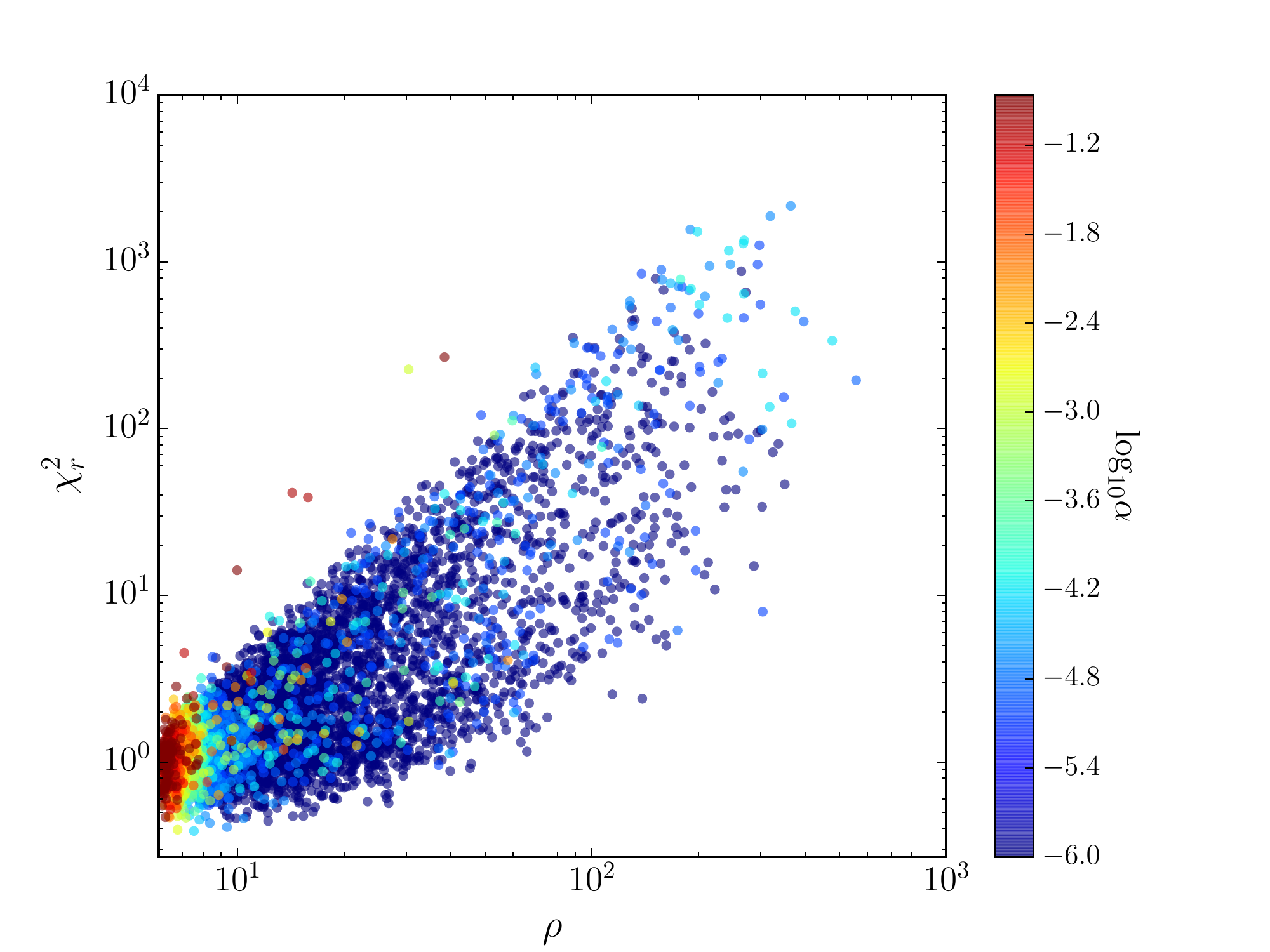}
\caption{Dependence of the RF classification for simulated signal triggers on $\chi^2_r$ (reduced 
chi-squared, expectation value $1$ in Gaussian noise) and SNR $\rho$ : color represents the 
false positive rate $\alpha$ at the threshold of the trigger's RF score.  We show illustrative results from LHO data for the RF classification with unit weights and minimum leaf size 8.}
\label{fig:H1_chisq_vs_snr_w1_l8}
\end{figure}
Figure~\ref{fig:H1_chisq_vs_snr_w1_l8} plots $\chi^2$ vs SNR for signal triggers and assigns 
to each trigger a color based on false positive rate $\alpha$, computed using the 
signal trigger's probability score (estimated by the RF) as a threshold. 

Most signal triggers are located below a diagonal boundary in the $\chi^2$-SNR plane. Signal 
triggers with low SNR are assigned higher $\alpha$ by the RF. The $\alpha$ values decrease by 
orders of magnitude with increasing SNR, for SNRs between $\sim 6$ and $\sim 10$. Between 
SNRs $\sim 10$ and $\sim 200$, the plot points maintain a low $\alpha$. Above SNR $\sim 200$, 
the $\alpha$ starts to increase again. The RF also assigns high $\alpha$ values to all 
signal triggers above the diagonal boundary and to a few triggers below the boundary.

The higher the false positive rate, the lower the probability score assigned by the RF 
classifier to the signal trigger. One would expect the RF to output a low probability score if 
a signal trigger is located in a region with a large number of noise triggers. Indeed, the 
region above the diagonal boundary, as well as the low- and high-SNR regions are populated by 
noise triggers. 

\begin{figure*}[htb]
\centering
\subfigure[\ No additional features]{%
\includegraphics[trim=10 5 15 15, clip, width=.32\linewidth]{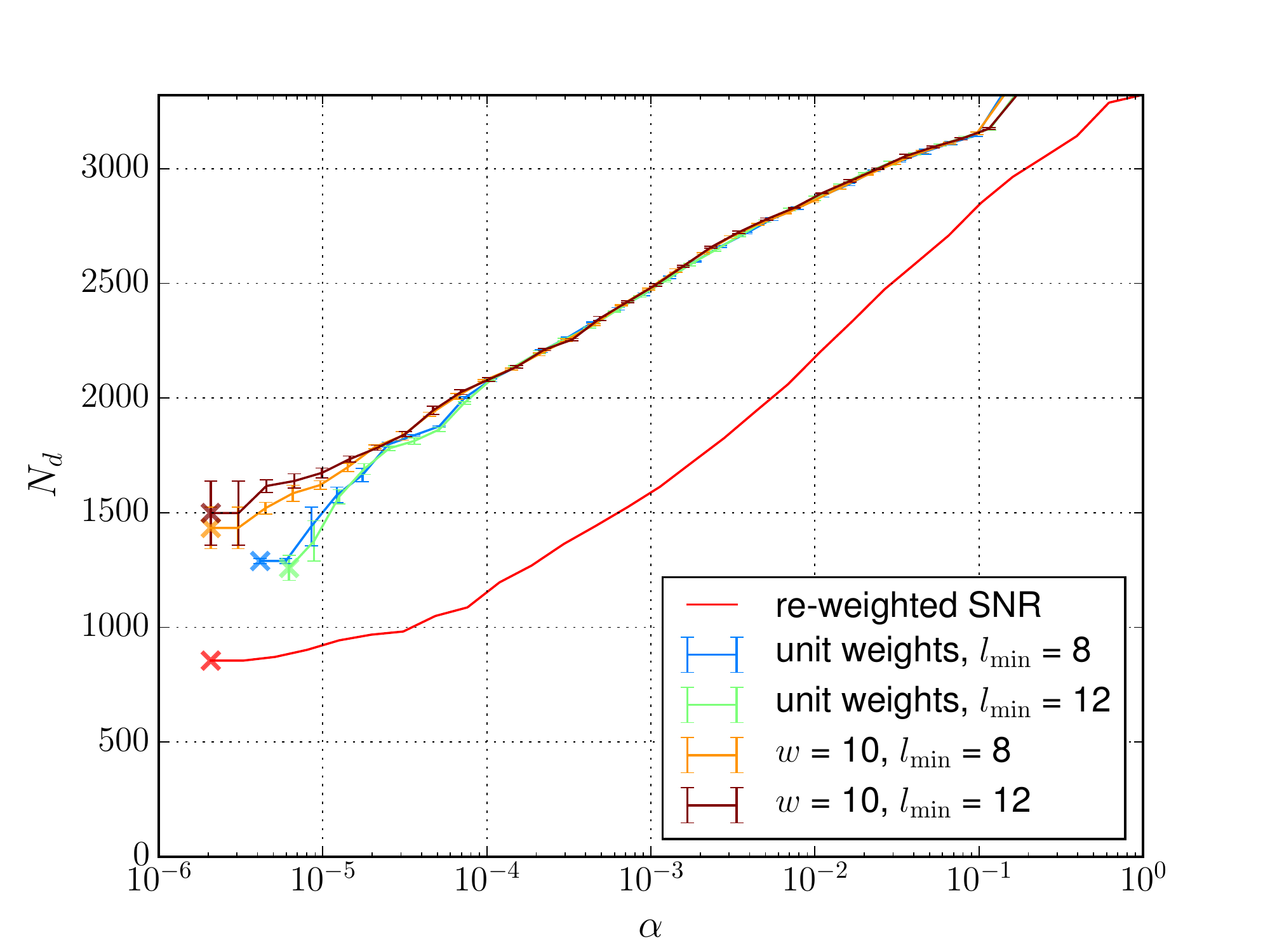}
\label{fig:H1_ROC_w1-10_l8-12}}
\subfigure[\ $\chi^2$ as an additional feature]{%
\includegraphics[trim=10 5 15 15, clip, width=.32\linewidth]{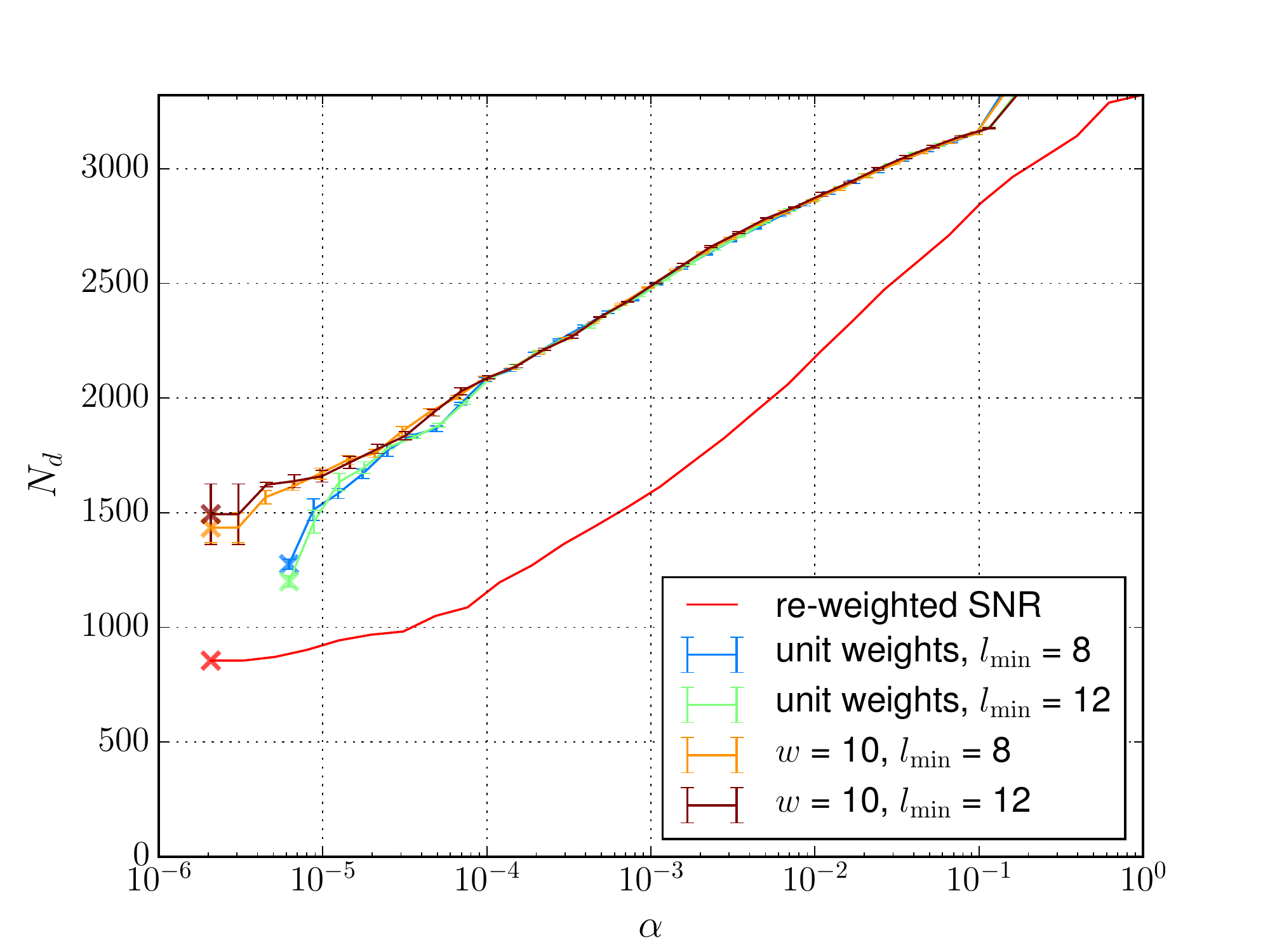}
\label{fig:H1_ROC_chisq_w1-10_l8-12}}
\subfigure[\ Re-weighted SNR as an additional feature]{%
\includegraphics[trim=10 5 15 15, clip, width=.32\linewidth]{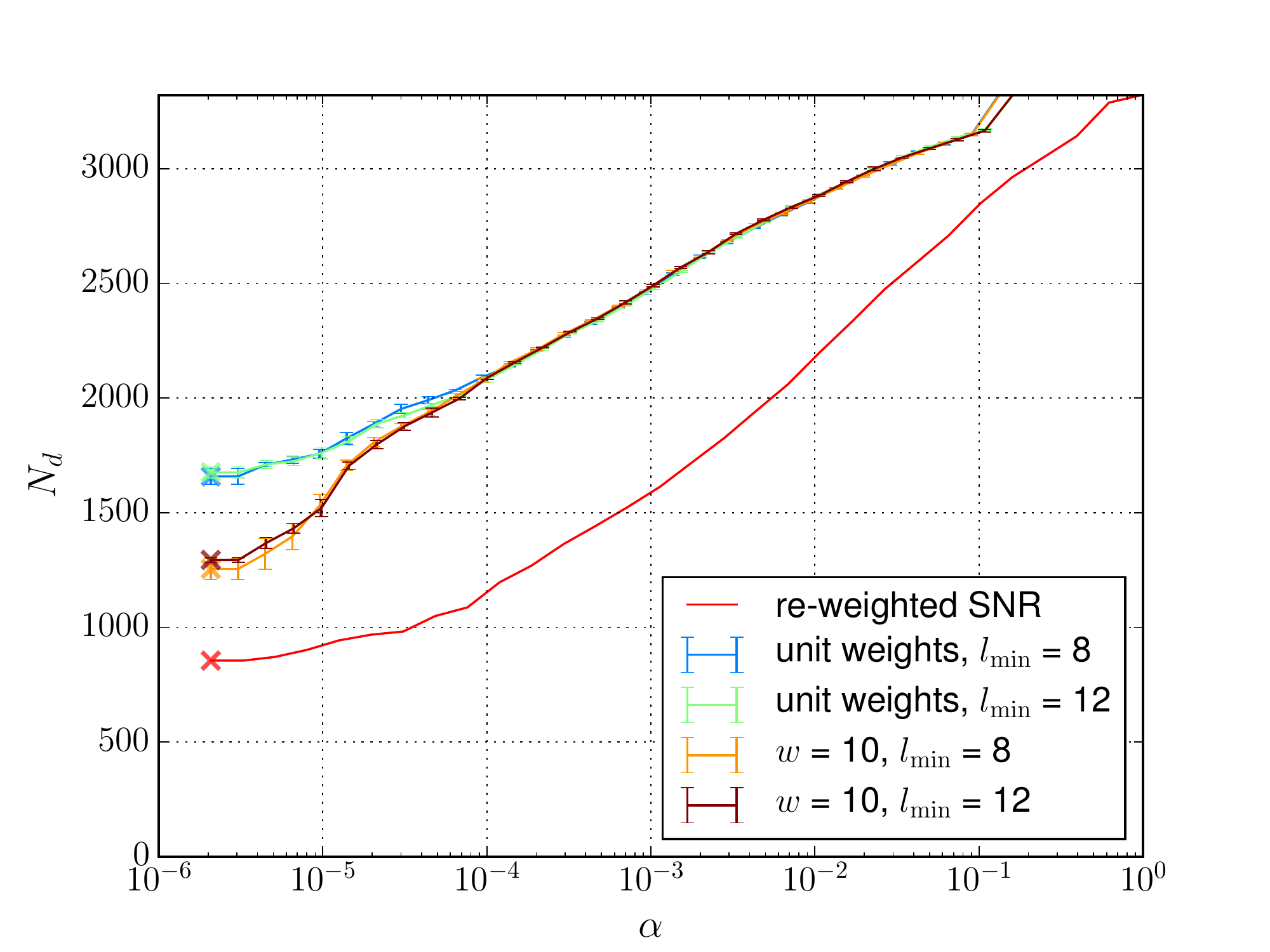}
\label{fig:H1_ROC_newsnr_1-10_l8-12}}
\caption{
ROC plots for the RF classifier comparing the standard feature set 
(left) to the cases where $\chi^2$ (centre) or re-weighted SNR $\hat{\rho}$ (right) are 
added to the feature sets.  Axes are as in figure~\ref{fig:H1_L1_ROC_plots}.  For each plot, the red 
line shows the ROC of the standard matched filter re-weighted SNR, while the colored lines with 
error bars show ROCs for the RF classifier with different minimum leaf sizes and weighting schemes.  
Including $\chi^2$ or $\hat{\rho}$ as an additional feature does not significantly increase
detection efficiency.  We show plots for LHO data only, similar results are obtained for LLO.}
\label{fig:H1_ROC_newsnr_chisq}
\end{figure*}

It is clear from Figure \ref{fig:H1_L1_ROC_plots} that the number of detected signals increases 
significantly when the RF is used to classify triggers, as compared to the standard re-weighted SNR 
$\hat{\rho}$ statistic. Na{\" \i}vely, one might think that using $\hat{\rho}$ as an additional 
feature for RF classification may further increase the detection efficiency. However, as seen in 
Figure~\ref{fig:H1_ROC_newsnr_chisq}, there is in fact no significant improvement by adding either 
$\chi^2$ or $\hat{\rho}$ as a feature.


\subsection{Classification with SNR and chi-squared only}

Our results show that a classifier using multiple features derived from triggers 
close to the time of a candidate event achieves a detection efficiency significantly higher than 
the standard re-weighted SNR statistic $\hat{\rho}$ via more accurate classification into signal
or noise.  An interesting question here is whether $\hat{\rho}$,
which is built from $\chi^2$ and SNR, is an optimal detection statistic \emph{given only the values 
of SNR and} $\chi^2$?  I.e.\ could a different function of (SNR,$\chi^2$) yield a higher
efficiency at fixed FPR? 
To address this question, we employ our RF classifier but use as features 
only the candidate triggers' SNR and $\chi^2$ values.  We find that that the detection efficiency 
given by the $\hat{\rho}$ statistic and the RF classifier, for a selection of different tree
parameters, are nearly equal over almost the entire range of false positive rate $\alpha$.  
(We also investigated nearest-neighbor and SVM classifiers in various configurations on this 
2-d problem and none produced any significant gain over the $\hat{\rho}$ statistic.)  
While not a proof, this is consistent with $\hat{\rho}$ being close to the optimum detection 
statistic that can be constructed from $\chi^2$ and SNR, at least for our data sets.  
%
\begin{figure*}[htb]
\centering
\subfigure[]{
  \includegraphics[trim=10 5 15 15, clip, width=.33\linewidth]{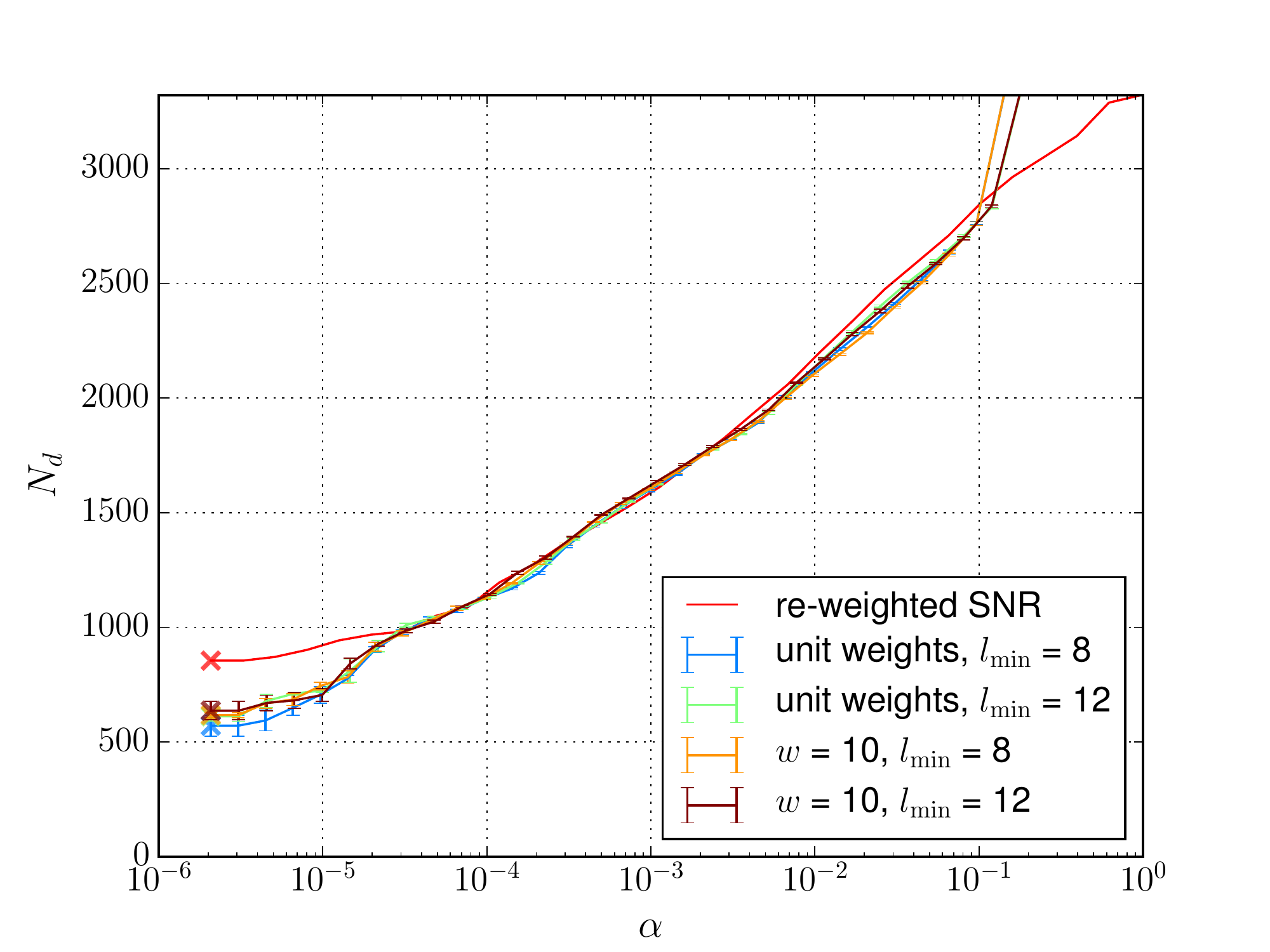}
  \label{H1-snr-chisq}}
\qquad
\subfigure[]{
  \includegraphics[trim=10 5 15 15, clip, width=.33\linewidth]{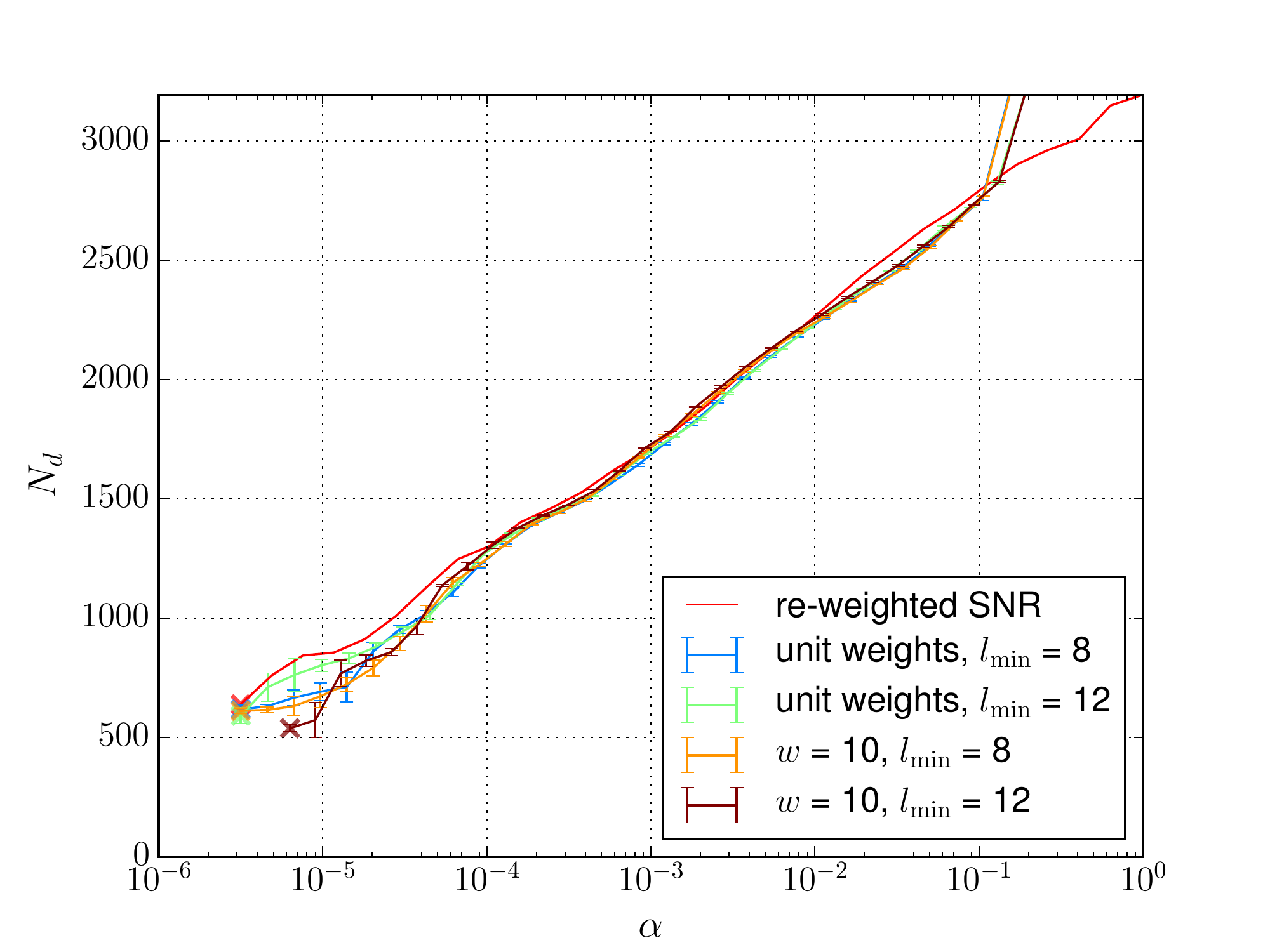}
  \label{L1-snr-chisq}}
\caption{ROC plots with LHO (left) and LLO (right) data comparing detection efficiency between the 
standard re-weighted SNR statistic (red lines) and the RF classifier using only 
the $\chi^2$ and SNR of individual triggers as features, for various choices of forest 
hyperparameters. 
The RF classifier reproduces the performance of the standard statistic over a wide range of false 
positive rate $\alpha$.}
\end{figure*}

\section{Discussion} \label{sec:concl}
\noindent 
In this study we have shown that a multivariate approach to classifying single-detector events as 
either (simulated) binary inspiral signals or transient noise artefacts (`glitches') can yield a 
more accurate classification than the currently used time-frequency chi-squared 
test~\cite{Brucechisq} in realistic early Advanced LIGO mock data; such an approach can thus be 
expected to result in a higher detection efficiency for inspiral signals at fixed false alarm 
probability. 

The chi-squared test, which uses only information present in the template matched filter integrand, 
can effectively downweight many high-amplitude glitches.  However, some transient artefacts with 
moderate SNR values still receive $\chi^2$ values typical of signals, although visual inspection of 
the data suggests a strong inconsistency.  These low-$\chi^2$ glitches are limiting on the 
sensitivity of the search to inspiral signals.  The use of other test quantities ('features') 
computed from data, which can contain information independent of the SNR and $\chi^2$ values,  
enables us to successfully suppress such glitches. 

We have conducted a detailed proof of principle study to identify and calculate various features 
from single-detector data, using both properties of the templated matched filter search (trigger 
count and maximum SNR over the whole template bank in time windows around a candidate event) and 
properties of triggers from a sine-Gaussian excess power search using the Q-transform.  We then set 
up a random forest classifier using sets of noise (glitch) triggers obtained from realistic 
simulated Advanced LIGO noise, and simulated signal triggers obtained from injecting NS-BH inspiral 
waveforms into the same noise data streams.  Note that the (orbitally precessing) injected signals 
do not lie in the space of (non-spinning) template waveforms; the resulting lack of effectualness 
is a proxy for unknown systematic uncertainties in NS-BH waveforms which may affect detection. 

The performance of the classifier was assessed by calculating the ROC for test data sets under 
two-fold cross-validation, and we estimated the variance of the ROC by running the cross-validation 
analysis several times with different random splits.  We also explored a range of different
hyperparameters in constructing the forest, notably the size of the `leaves' and the relative 
weighting of signal and noise events. 

The figure of merit for the classification is the expected number of signals found at fixed false 
positive rate (FPR), which is proportional to the volume of space that the search would be 
sensitive to, assuming that signal events are uniformly distributed through space.  We find an 
increase of a factor 1.5--2 in this sensitive volume at low FPR compared to the ``$\chi^2$-
reweighted SNR'' statistic currently used in LIGO searches. 
The improvement in sensitivity is relatively insensitive to different choices of forest parameters, 
and is not substantially affected by including or omitting the $\chi^2$ or re-weighted SNR values as 
features.  
We also checked that, given only the single-detector SNR and $\chi^2$ values for individual 
triggers, the classifier was able to reproduce the ROC of the reweighted SNR statistic; thus, any 
increase in sensitivity for the full classifier is due to the inclusion of additional information.

This is the first demonstration that a multivariate machine learning classifier acting on 
\emph{single-detector} data can outperform the reweighted SNR statistic for inspiral signals.  
Although a welcome proof of principle, various issues remain to be addressed before such a 
classifier may be used in production analysis of Advanced LIGO-Virgo data.  

The trigger set used here was restricted to relatively high matched filter SNR ($\rho > 6.0$) and 
was clustered over the template bank (selection of the highest-SNR trigger inside given time 
windows), in order to obtain a manageable number of samples for training the random forest (few$\,
\times 10^5$); typically under an hour was required for each training round on a single CPU.  
In current inspiral searches, however, a threshold $\rho > 5.5$ is used and no clustering over the 
bank is applied~\cite{CBCCompanion,Usman} thus the total rate of triggers is some orders of 
magnitude higher.  Computational cost might then be addressed by parallelizing the training
stage over hundreds of independent trees in the forest, by a decimation process to select a small 
subset of triggers for training, or possibly use of GPUs. 

Since our multivariate classifier is able to efficiently sort signal from noise events in a single 
detector, we expect that its use within a coincident (two- or more-detector) search would also
increase sensitivity to GW inspiral signals.  A detection statistic for coincident events 
(i.e.\ sets of triggers having consistent parameters over the network) would incorporate the 
single-detector $\hat{p}$ scores in addition to coincidence parameters: time delays, phase 
differences and relative amplitudes between detectors~\cite{Nitz2017}.  A more direct application 
of the method concerns signals which may arrive when only a single detector is 
observing~\cite{Callister:2017urp} as expected in ground-based detector networks with typical duty 
cycles well below 100$\%$.  Although the 
significance of single-detector candidates is limited by the amount of data available, signals might 
still be identified if their expected rate, given the detector sensitivity, is high enough and if a 
reliable method of distinguishing them from noise artefacts is available.

A natural extension of our methodology would be to other types of compact binary coalescence signal 
templates, specifically for higher-mass systems, given that the traditional $\chi^2$ test is less 
effective for such systems, having shorter duration templates~\cite{CBCCompanion,Nitz2017}.  An
expansion of the method to the early Advanced LIGO stellar-mass binary search space 
of~\cite{CBCCompanion} (template binary mass in $2$-$100$\,M$_\odot$, non-precessing component spins 
with dimensionless magnitudes up to $0.9895$) is under development for integration in the PyCBC
pipeline (for which see \url{https://ligo-cbc.github.io/} and \cite{Usman,Canton:2014ena}). 

Finally, since many transient artefacts in the gravitational wave strain channel could in principle
be predicted by use of auxiliary channels that monitor the state of the instrument and the 
environment~\cite{DetcharCompanion,S5DQ,S6DC}, we could also consider deriving input features from 
these channels.  This implies a significant increase in the dimensionality and volume of data input 
to the classifier, requiring machine learning techniques robust to high-dimensional multivariate 
data where many channels may be redundant or contain their own, irrelevant artefacts 
(see~\cite{Biswas:2013wfa} for applications of machine learning to noise artefacts in Initial 
LIGO data).

\section*{Acknowledgements} \noindent
We would like to thank Florent Robinet for providing and maintaining the Omicron pipeline used 
for this study; Andrew Lundgren and Badri Krishnan for enlightening discussions; 
the LIGO Scientific Collaboration, in particular the Compact Binary Coalescence working group, for 
the recoloured Early Advanced LIGO mock data set used in this study; and the 
Albert-Einstein-Institut (Hannover) for the use of the Atlas computing cluster. 

SJK is grateful for the hospitality and travel support received from the Max-Planck-Gesellschaft; he 
further acknowledges travel support from the Raymond Hughes Foundation at the University of Arkansas. 
TD acknowledges support from the Max-Planck-Gesellschaft. TDC was supported by
the International Max-Planck Research School on Gravitational-Wave Astronomy, and 
by an appointment to the NASA Postdoctoral Program at the Goddard Space Flight Center, 
administered by Universities Space Research Association under contract with NASA.



%

\end{document}